\RequirePackage{lineno}
\documentclass[prd,twoline,twocolumn,amsmath,amssymb,showpacs]{revtex4}  % for review and submission 
\usepackage{graphicx}% Include figure files
\usepackage{dcolumn}% Align table columns on decimal point
\usepackage{bm}% bold math
\usepackage{amssymb,amsmath}
\usepackage{subfigure}
\usepackage{verbatim}
\begin{document}

\mathchardef\mhyphen="2D
\newcommand{\dzero}     {D0}
\newcommand{\ttbar}     {\mbox{$t\bar{t}$}}
\newcommand{\bbbar}     {\mbox{$b\bar{b}$}}
\newcommand{\ccbar}     {\mbox{$c\bar{c}$}}
\newcommand{\ppbar}     {\mbox{$p\bar{p}$}}
\newcommand{\herwig}    {{\sc herwig}}
\newcommand{\pythia}    {{\sc pythia}}
\newcommand{\alpgen}    {{\sc alpgen}} 
\newcommand{\qq}        {{\sc qq}} 
\newcommand{\evtgen}    {{\sc evtgen}} 
\newcommand{\tauola}    {{\sc tauola}} 
\newcommand{\geant}     {{\sc geant}}
\newcommand{\cteq}	{CTEQ6L1}
\newcommand{\minuit}	{{\sc minuit}}
\newcommand{\met}       {\mbox{$\not\!\!E_T$}}
\newcommand{\mex}       {\mbox{$\not\!\!E_x$}}
\newcommand{\mey}       {\mbox{$\not\!\!E_y$}}
\newcommand{\metcal}    {\mbox{$\not\!\!E_{Tcal}$}}
\newcommand{\metzfit}   {\mbox{$\not\!\!E_{T}^{Z\mhyphen \rm \/ fit}$}}
\newcommand{\metjes}	{\mbox{$\not\!\!E_{T}^{\rm \/ JES}$}}
\newcommand{\mexunc}	{\mbox{$\not\!\!E_{x}^{\rm \/ u}$}}
\newcommand{\meyunc}	{\mbox{$\not\!\!E_{y}^{\rm \/ u}$}}
\newcommand{\meiunc}	{\mbox{$\not\!\!E_{i}^{\rm \/ u}$}}
\newcommand{\meicalc}   {\mbox{$\not\!\!E_{i}^{\rm \/ calc}$}}
\newcommand{\meiobs}    {\mbox{$\not\!\!E_{i}^{\rm \/ obs}$}}
\newcommand{\rar}       {\rightarrow}
\newcommand{\eps}       {\epsilon}
\newcommand{\pt}        {$p_T$}
\newcommand{\approxle}  {\mathrel{\vcenter{\hbox{$\buildrel<\over\sim$}}}}
\newcommand{\nuwt}	{$\nu$WT}
\newcommand{\nuwth}	{$\nu$WT$_h$}
\newcommand{\nuwtf}	{$\nu$WT$_f$}
\newcommand{\etal}	{\textit{et al.}}
\newcommand{\ltrk}	{$\ell+$track}
\newcommand{\etrk}	{$e+$track}
\newcommand{\mutrk}	{$\mu+$track}
\newcommand{\mtfit}	{$m_t^{\rm fit}$}
\newcommand{\mlt}	{$M_{\ell \rm t}$}

%%%%%%%%%%%%%%%%%%%%% Publisher's Area please ignore %%%%%%%%%%%%%%%
%
%\catchline{}{}{}{}{}
%
%%%%%%%%%%%%%%%%%%%%%%%%%%%%%%%%%%%%%%%%%%%%%%%%%%%%%%%%%%%%%%%%%%%%
%\linenumbers

\hspace{5.2in} \mbox{FERMILAB-PUB-09/125-E}

\title{Measurement of the top quark mass in final states with two leptons}
% LIST_OF_AUTHORS_R2.TEX                 3/27/09            
%
\author{V.M.~Abazov$^{37}$}
\author{B.~Abbott$^{75}$}
\author{M.~Abolins$^{65}$}
\author{B.S.~Acharya$^{30}$}
\author{M.~Adams$^{51}$}
\author{T.~Adams$^{49}$}
\author{E.~Aguilo$^{6}$}
\author{M.~Ahsan$^{59}$}
\author{G.D.~Alexeev$^{37}$}
\author{G.~Alkhazov$^{41}$}
% ...

\date{April 20, 2009}

%\maketitle

%\begin{history}
%\received{\today}
%\revised{Day Month Year}
%\end{history}
\begin{linenomath}
\begin{abstract}
\vspace*{3.0cm}
We present
measurements of the top quark mass ($m_t$) in \ttbar\ candidate events with
two final state leptons using
1 fb$^{-1}$ of data collected by the \dzero\ experiment.  Our data
sample is selected by requiring two fully identified
leptons or by relaxing one lepton requirement 
to an isolated track if at least one jet is tagged as a $b$ jet.
The top quark mass is extracted after reconstructing the event kinematics 
under the \ttbar\ hypothesis using two methods.  In the first method, we 
integrate over
expected neutrino rapidity distributions, and in the second we 
calculate a weight for the possible top quark masses based on the 
observed particle momenta and the known parton distribution
functions.  We analyze 83 candidate events in data and obtain
$m_t=176.2\pm4.8 (\rm stat) \pm 2.1 (\rm sys)$~GeV
and $m_t=173.2\pm 4.9 (\rm stat) \pm 2.0 (\rm sys)$~GeV for the
two methods, respectively.
Accounting for correlations between the two methods, we combine the
measurements to obtain $m_t=174.7 \pm 4.4 (\rm stat) \pm 2.0 (sys)$~GeV.
\end{abstract}
\end{linenomath}
\noindent \keywords{top quark; Yukawa coupling; fermion mass.}
\newline
\pacs{12.15.Ff, 14.65.Ha}

\maketitle

\section{Introduction}
After the top quark was discovered in 1995 \cite{d0Obs,cdfObs}, 
emphasis quickly turned
to detailed studies of its properties including 
measuring its mass across all reconstructable final states. 
Within the standard model, 
a precise measurement of the top quark mass ($m_t$) and $W$ boson mass ($M_W$) 
can be used to constrain the Higgs boson mass ($M_H$).
In fact, these masses  can be related by radiative corrections
to $M_W$.  One-loop corrections give
$M_{W}^{2}=\frac{\pi\alpha/\sqrt{2}G_{F}}{{\rm sin}^2\theta_{W}(1-\Delta r)}$,
where $\Delta r$ depends quadratically on $m_t$ and logarithmically on 
$M_H$~\cite{mtrole}.
Beyond its relation to $M_H$, the top quark mass reflects
the Yukawa coupling, $Y_t$, for the top quark via $Y_t = m_t \sqrt{2}/v$,
where $v=246$~GeV is the vacuum expectation value of the Higgs field \cite{PDG}.
Given that these couplings are not predicted by the theory, $Y_t=0.995\pm0.007$ 
for the current $m_t$~\cite{WA} is curiously close to unity.  
One of several possible modifications to the mechanism
underlying electroweak symmetry breaking suggests a more central role for the top quark.
For instance, in top-color assisted technicolor \cite{topcolor1,topcolor2}, the 
top quark plays a major role in electroweak symmetry breaking.
These models entirely remove the need for an elementary
scalar Higgs field in favor of new strong interactions that provide the observed 
mass spectrum.  Perhaps there are extra Higgs doublets as in MSSM 
models~\cite{mssm}; measurement of the top quark mass may be sensitive to such models
(e.g., Ref.~\cite{mssm1}).

In the standard model, $BR(t\rightarrow Wb)$ is expected to be nearly 100\%.  
So the relative rates of final states in events with top quark
pairs, \ttbar, 
are dictated by the branching ratios of the $W$ boson to various fermion pairs.
In approximately 10\% of \ttbar\ events, both $W$ bosons decay
leptonically.  
Generally, only events that include the $W\to e\nu$ and $W\to\mu\nu$ modes 
yield final states with precisely reconstructed lepton momenta to be used for 
mass analysis.  Thus, analyzable dilepton final states are
\ttbar~$\to\ell\bar{\ell}'+\bar{\nu}\nu'+\bbbar$, where $\ell, \ell'=e,\mu$.  
We measure $m_t$ in these dilepton events.
The $W \to \tau\nu \to e(\mu)\nu\bar{\nu}$ decay modes cannot be
separated from the direct $W \to e(\mu)\nu$ decays and are included in our analysis.

Dilepton channels provide a sample that is statistically independent
of the more copious \ttbar~$\to \ell\nu+q\bar{q'}+b\bar{b}$ ($\ell+$jets)
decays.  The relative contributions of specific systematic effects
are somewhat different between mass measurements from events with
dilepton or 
$\ell+$jets final states.
The jet multiplicity and the 
dominant background processes are different.  The measurement of $m_t$ in
the dilepton channel also
provides a consistency test of the \ttbar\ event
sample with the expected $t\rightarrow Wb$ decay.  Non-standard
decays of the top quark, such as $t\rightarrow H^{\pm}b$, can
affect the final state particle kinematics differently in different
\ttbar\ channels.  These kinematics affect the
reconstructed mass significantly, for example in
the $\ell+$jets channel~\cite{chHiggs}.  
Therefore, it is important to precisely test the consistency of the $m_t$
measurements in different channels.

Previous efforts to measure $m_t$ in the dilepton channels have been pursued by the
\dzero\ and CDF collaborations.  A frequently used technique reconstructs 
individual event kinematics using known constraints
to obtain a relative probability of 
consistency with a range of top quark masses.
The ``matrix weighting" method (MWT) 
follows the ideas proposed by Dalitz and Goldstein
\cite{dalitz} and Kondo \cite{kondo}.  It
uses parton distribution functions
and observed particle momenta to obtain a mass estimate for each
dilepton event, and has previously been implemented
by \dzero\ ~\cite{r1d0mt,r2d0mt}.
The ``neutrino weighting" method (\nuwt) was developed at \dzero ~\cite{r1d0mt}.
It integrates over expected neutrino rapidity
distributions, and has been used by
both the \dzero\ ~\cite{r1d0mt,r2d0mt} and CDF~\cite{r2CDFnuWT} collaborations.

In this paper, we describe a measurement of the top quark mass in 1 fb$^{-1}$
of \ppbar\ collider data collected using the \dzero\ detector at the Fermilab
Tevatron Collider.  Events are selected in two 
categories.
Those with one fully identified electron and one fully identified muon, two 
electrons, or two muons are referred to as ``$2\ell$."  To
improve acceptance, we include a second category consisting of 
events with only one fully reconstructed electron 
or muon and an isolated high transverse momentum ($p_T$) 
track as well as at least one identified $b$ jet, which we refer to 
as ``\ltrk" events.  We
describe the detection, selection, and modeling of these events in
Sections~\ref{sec:apparat} and~\ref{sec:evtSel}.
Reconstruction of the kinematics of \ttbar\ events proceeds by both the MWT
and \nuwt\ approaches.  These methods are described
in Section ~\ref{sec:kinRec}.  
In Section~\ref{sec:LikeHood}, we describe the maximum likelihood
fits to extract $m_t$ from data.
Finally, we discuss our results and systematic uncertainties in 
Section~\ref{sec:Results}.

{\section{Detector and Data Sample}
\label{sec:apparat}}
\subsection{Detector Components}
The \dzero\ Run II detector \cite{d0nim} is a multipurpose collider detector
consisting of an inner magnetic central tracking system,
calorimeters, and outer muon tracking detectors.
The spatial coordinates of the \dzero\ detector are defined as
follows: the positive $z$ axis is along the direction of the
proton beam while positive $y$ is defined as upward from
the detector's center, which serves as the origin.
The polar angle $\theta$ is measured with respect to the positive $z$
direction and is usually expressed as
the pseudorapidity, $\eta\equiv-\ln[\tan(\theta/2)]$.
The azimuthal angle $\phi$ is measured with
respect to the positive $x$ direction, which points away from the center of the 
Tevatron ring.

The inner tracking detectors are responsible for measuring the trajectories 
and momenta of charged particles and for locating track vertices. 
They reside inside a superconducting solenoid that generates a magnetic
field of 2 T.  A silicon microstrip tracker (SMT) is innermost and 
provides precision position measurements,
particularly in the azimuthal plane, which allow the reconstruction of
displaced secondary vertices from the decay of long-lived particles.  This 
permits identification of jets from heavy flavor quarks, particularly $b$ 
quarks.  A central fiber tracker
is composed of scintillating fibers mounted on eight concentric 
support cylinders.
Each cylinder supports one axial and one stereo layer of fibers, 
alternating by  $\pm3\,^{\circ}$ relative to the cylinder axis.
The outermost cylinder provides coverage for $|\eta|<1.7$.

The calorimeter measures electron and jet energies, directions, and shower
shapes relevant for particle identification.  Neutrinos are also
measured via the calorimeters' hermeticity and the constraint of momentum
conservation in the plane transverse to the beam direction.
Three liquid-argon-filled cryostats containing primarily uranium absorbers constitute the
central and endcap calorimeter systems.
The former covers $|\eta| < 1.1$, and the latter 
extends coverage to $|\eta|= 4.2$. Each calorimeter
consists of an electromagnetic (EM) section followed longitudinally
by hadronic sections.  Readout cells are arranged
in a pseudo-projective geometry with respect to the nominal interaction region.

Drift tubes and scintillators are arranged in planes outside
the calorimeter system to measure
the trajectories of penetrating muons.  One drift tube layer resides
inside iron toroids with a magnetic field of 1.8 T, while two more layers are 
located outside.  The coverage of the muon system is $|\eta| < 2$.

\subsection{Data Sample}

The \dzero\ trigger and data acquisition systems are designed to
accommodate instantaneous luminosities up to $3\times 10^{32}$~$\rm cm^{-2}s^{-1}$. 
The Tevatron operates with 396~ns spacing between proton (antiproton)
bunches and delivers a 2~MHz bunch crossing rate.  For our data
sample, each crossing
yields on average 1.2 \ppbar\ interactions. 
%in addition to the hard scatter.

Luminosity measurement at \dzero\ is based on the rate of
inelastic $p\bar{p}$ collisions observed by 
plastic scintillation counters mounted on the inner
faces of the calorimeter endcap cryostats.  Based on
information from the tracking, calorimeter, and
muon systems, the first level of the trigger
limits the rate for accepted events to
2~kHz. This is a dedicated hardware trigger.
Second and third level triggers employ algorithms running in processors to 
reduce the output rate to about 100~Hz, which is written to tape.

Several different triggers are used for the five decay channels considered
in this measurement.  We employ single electron triggers
for the $ee$ and $e+$track channels and
single muon triggers for the $\mu\mu$ and $\mu+$track
channels.  The $e\mu$ analysis employs all unprescaled triggers requiring
one electron and/or one muon.  We also use triggers requiring one lepton plus
one jet for the \ltrk\ channels.
A slight difference between the \nuwt\ and MWT
analyses occurs because the latter excludes 2\%
of data collected while the single muon trigger was prescaled.  While the effect on
the kinematic distributions is negligible, this results in one less $\mu\mu$
candidate event in the final sample for the MWT analysis.

Events were collected with these
triggers at \dzero\ between April 2002 and
February 2006 with $\sqrt s=1.96$~TeV.  Data quality requirements
remove events for which the tracker, calorimeter, or muon system are 
known to be functioning improperly.  The integrated luminosity of the
analyzed data sample is about 1 fb$^{-1}$.

{\subsection{Particle Identification}
\label{sec:pid}}
We reconstruct the recorded data to identify and
measure final state particles, as described below.
A more detailed description can be found
in Ref.~\cite{d0ll425pbcsec}.

The primary event vertex (PV) is identified as the vertex with the lowest
probability to come from a soft \ppbar\ interaction based on the transverse momenta of
associated tracks.  We select events in which the PV is reconstructed from at least 
three tracks and with $|z_{PV}|<$ 60 cm.  Secondary vertices from the decay of 
long-lived particles from the hard interaction are reconstructed from two or more tracks
satisfying the requirements of $p_T>1$~GeV and more than one hit in the SMT.  
We require each track to have a large impact parameter significance,
$\rm DCA/\sigma_{DCA} > 3.5$, with respect to the PV, where DCA is the
distance of the track's closest approach to the PV in the transverse plane.

High-$p_T$ muons are identified by matching tracks in the inner tracker
with those in the muon system.  The track requirements include a cut on
DCA$<0.02$ (0.2) cm for tracks with (without) SMT hits. 
Muons are isolated in the tracker when
the sum of track momenta in a cone of radius
$\Delta {\mathcal R}({\rm muon,track}) = \sqrt{(\Delta\eta)^2+(\Delta\phi)^2}=0.5$
around the muon's matching track is small compared to the track \pt.
We also require isolated muons to have 
the sum of calorimeter cell energies in an annulus with radius in the range 
$0.1<\Delta {\mathcal R}<0.4$ around the 
matched track to be low compared to the matching track~\pt.  

High-$p_T$ isolated tracks are identified solely in the inner
tracker.  We require them to satisfy track isolation requirements and 
to be separated from calorimeter jets by $\Delta {\mathcal R}(\rm jet,track)>0.5$.
These tracks must correspond to leptons from the PV, so we 
also require that $\rm DCA/\sigma_{DCA}<2.5$.
We avoid double-counting leptons
by requiring $\Delta {\mathcal R}(\rm track, \ell) > 0.5$.

Electrons are identified in the EM calorimeter.  Cells are clustered 
according to a cone algorithm within
$\Delta {\mathcal R} < 0.2$ and then matched with an inner detector track. 
Electron candidates are required to deposit 90\% of their energy in the
EM section of the calorimeter.  They must also
satisfy an initial selection which includes a shower shape
test ($\chi^2_{\rm\/hmx}$) with respect to the expected electron shower shape,
and a calorimeter isolation requirement summing 
calorimeter energy within
$\Delta {\mathcal R}<0.4$ but excluding the cluster energy.  To further
remove backgrounds, a likelihood ($\mathcal L_e$) selection is 
determined based on seven tracking and calorimeter parameters, including
$\chi^2_{\rm\/hmx}$, DCA, and track isolation calculated in an annulus of
$0.05<\Delta {\mathcal R} <0.4$ around the electron. 
The final electron energy calibration is 
determined by comparing the invariant mass of high $p_T$ 
electron pairs in $Z/\gamma^*\to e^+e^-$ events with the
world average value of the $Z$ boson mass as measured by the LEP experiments
\cite{PDG}. 

In \ttbar\ events, the leptons and tracks originate from the hard interaction.
Therefore, we require their $z$ positions at the closest approach
to the beam axis to match that of the PV within 1 cm. 

We reconstruct jets using a fixed cone algorithm \cite{conealg} with radius of $0.5$.
The four-momentum
of a jet is measured as the sum of the four-momenta assigned to calorimeter cells
inside of this cone.  
We select jets that have a longitudinal shower profile consistent with
that of a collection of charged and neutral hadrons.  We confirm jets via the
electronically independent calorimeter trigger readout chain.
Jets from $b$ quarks are tagged using a neural network $b$ jet tagging
algorithm \cite{d0btag}.  This combines the impact parameters for all 
tracks in a jet, as well as information about reconstructed secondary 
vertices in the jet.  
We obtain a typical efficiency of 54\% for $b$ jets
with $|\eta| < 2.4$ and $p_T>30$~GeV for a selection which accepts 
only 1\% of light flavor ($u$, $d$, $s$ quark or gluon) jets.

Because the $b$ jets carry away much of the rest energy of the top quarks,
it is critical for the measurement of $m_t$ that the measurements of the
energies of jets from top quark decay be well calibrated.
Jet energies determined from the initial cell energies do not 
correspond to the energies of final state particles striking the calorimeter.
As a result, a detailed calibration is applied~\cite{d0r2jinc,d0jes} in data and
Monte Carlo separately.  In
general, the energy of all final state particles inside the jet cone, $E_j^{\rm \/ptcl}$,
can be related to the energy measured inside the jet cone, $E_j$, by
$E_j^{\rm ptcl}$ $=(E_j-O)/( R\,S )$.
Here, $O$ denotes an offset energy primarily from extra interactions
in or out of time with an event.
$R$ is the cumulative response of the calorimeter to all of the particles in a
jet.  $S$ is the net energy loss due to
showering out of the jet cone.  For a given cone radius, $O$ and $S$ are
functions of the jet $\eta$ within the detector.  $O$ is also a
function of the number of reconstructed event vertices and the
instantaneous luminosity.  $R$ is the largest correction and 
reflects the lower response
of the calorimeters to charged hadrons relative to electrons and photons.  It
also includes the effect of energy losses in front of the calorimeter.  The 
primary response correction is derived $in$ $situ$ from 
$\gamma+$jet events and has
substantial dependences on jet energy and $\eta$.  For all jets that contain
a non-isolated muon, we add the muon momenta to that of the jet.  Under
the assumption that these are $b$ quark semileptonic decays, we also add 
an estimated 
average neutrino momentum assumed to be collinear with the jet direction.
The correction procedure discussed above does not correct all
the way back to the original $b$ quark parton energy.

The event missing transverse energy, \met, is equal in magnitude
and opposite in direction to the vector sum of all significant
transverse energies measured by the individual calorimeter cells.
It is corrected for the transverse momenta of all 
isolated muons, as well as for the corrections to the electron and jet energies.
In the \ltrk\ channels, the \met\ is also corrected if the track does not point 
to a jet, electron, or muon.  In this case, we 
substitute the track $p_T$ for the calorimeter energy within a cone of radius 
$\Delta {\mathcal R}=0.4$ around the track.
\newline

{\subsection{Signal and Background Simulation}
\label{sec:mc}}

An accurate description of the composition and kinematic properties of the
selected data sample is essential to the mass measurement.  
Monte Carlo samples for the \ttbar\ processes are generated for several test values of 
the top quark mass. The event generation, 
fragmentation, and decay are performed by \pythia\ 6.319~\cite{pythia}.
Background processes are called ``physics" backgrounds when charged leptons 
originate from $W$ or $Z$ boson decay and when \met\ comes
from high $p_T$ neutrinos. Physics backgrounds include 
$Z/\gamma^{*}\to\tau\tau$ with $\tau\to e,\mu$ and diboson ($WW$, $WZ$, and $ZZ$)
production. The $Z/\gamma^{*}\to\tau\tau$ background processes are generated with
\alpgen\ 2.11 ~\cite{alpgen} as the event 
generator and \pythia\ for fragmentation and 
decay.  We decay hadrons with $b$ quarks using \evtgen ~\cite{evtgen}.
To avoid double counting QCD radiation between \alpgen\ and \pythia, 
the jet-parton matching scheme of Ref.~\cite{mlm} is employed in \alpgen. 
The diboson backgrounds are simulated with \pythia. We use 
the \cteq~\cite{cteq} parton distribution function (PDF).
Monte Carlo events are then processed through a \geant-based \cite{geant} 
simulation of the D0 detector. In order to accurately model the effects of multiple 
proton interactions and detector noise, data events from random $p\bar{p}$ crossings 
are overlaid on the Monte Carlo events.
Finally, Monte Carlo events are processed through the same reconstruction software as used for data.

In order to ensure that reconstructed objects in these samples reflect the performance
of the detector in data, several corrections are applied.  Monte Carlo events are
reweighted 
by the $z$ coordinate of the PV to match the profile in data. 
The Monte Carlo events are further tuned such that the efficiencies to find leptons, 
isolated tracks, and jets in Monte Carlo events match those determined from data.
These corrections depend on the $p_T$ and $\eta$ of these objects.  
The jet energy calibration derived for data is applied to jets in data, and the 
jet energy calibration derived for simulated events is applied to simulated events.
We observe a residual discrepancy
between jet energies in $Z+$jets events in data and Monte Carlo.  We 
apply an additional correction to jet energies in the Monte Carlo to bring
them into agreement with the data.
This adjustment is then propagated into the \met.
We apply additional smearing to the reconstructed jet and lepton transverse momenta
so that the object resolutions in Monte Carlo match those in data.
Owing to differences in $b$-tagging efficiency between data and simulation, 
$b$-tagging in Monte Carlo events is modeled
by assigning to each simulated event a weight defined as
\begin{linenomath}
\begin{equation}
P=1-\displaystyle\prod_{i=1}^{N_{\rm \/ jets}}[1-p_i(\eta, p_T, \rm flavor)],
\end{equation}
\end{linenomath}
where $p_i(\eta, p_T, \rm flavor)$ is the probability of the $i^{\rm \/ th}$ jet 
to be identified as originating
from a $b$ quark, obtained from data measurements.  This product is taken over all jets. 
Instrumental backgrounds are modeled from a combination of data and simulation
and are discussed in Section~\ref{sec:instrBG}.
\newline

{\section{Selected Event Sample}
\label{sec:evtSel}}

Events are selected for all channels by requiring either two leptons ($2\ell$) or a lepton
and an isolated track ($\ell+$track), each with
$p_T>15$ GeV.
Electrons must be within $\left|\eta\right|<1.1$ 
or $1.5<\left|\eta\right|<2.5$; muons and tracks should have 
$\left|\eta\right|<2.0$.  An opposite charge requirement is applied to the two
leptons or to the lepton and track.
At least two jets are also required with
pseudorapidity $\left|\eta\right|<2.5$ and $p_T>20$ 
GeV.  We require the leading jet to have $p_T>30$ GeV.
Since neutrinos coming from $W$ boson decays in $t\bar{t}$ events are a source of 
significant missing energy, a cut on \met\ is a powerful 
discriminant 
against background processes without neutrinos such as $Z/\gamma^*\to 
ee$ and $Z/\gamma^*\to \mu\mu$. 
All channels except $e\mu$ require at least \met\ $> 25$~GeV.
\newline

\subsection{$2\ell$ Selection}

Our selection of $2\ell$ events follows Ref.~\cite{d0ll1fbcsec}.
In the $ee$ channel, events with a dielectron invariant 
mass $M_{ee}<15$ GeV or $84<M_{ee}<100$ GeV are rejected.
We require \met\ $>35$ GeV and \met\ $>45$ GeV 
when $M_{ee}>100$ GeV and $15<M_{ee}<84$ GeV, respectively. 
In the $\mu\mu$ channel, we select events with $M_{\mu\mu}>30$ GeV and
\met\ $>40$ GeV.  To further reject the $Z/\gamma^{\ast}\rightarrow \mu\mu$ 
background in the $\mu\mu$ channel, we require that
the observed \met\ be inconsistent with arising solely from the 
resolutions of the measured muon momenta and jet energies.

In the $e\mu$ analysis, no cut on \met\ is applied because the 
main background process $Z/\gamma^*\to\tau\tau$ generates
four neutrinos having moderate $p_T$.
Instead, the final selection in this channel requires
$H^{\ell}_T=p^{\ell_1}_T+\sum(E^j_T)>115$~GeV, where
$p^{\ell_1}_T$ denotes the transverse momentum of the leading lepton, 
and the sum is performed over the two leading jets.
This requirement rejects the largest backgrounds for this
final state, $Z/\gamma^{\ast} \rightarrow \tau \tau$
and diboson production.  
We require the leading jet to have $p_T>40$ GeV.

The selection described above is derived from that used for the $t\bar{t}$ 
cross-section analysis.
Varying the \met\ and jet $p_T$ selections indicated that this selection 
minimizes the statistical uncertainty on the $m_t$ measurement.
We select 17 events in the $ee$ channel and 13 events
(12 events for MWT) in the $\mu\mu$ channel.
We select 39 events in the $e\mu$ channel.

\subsection{\ltrk\ Selection}

The selection for the \ltrk\ channels is similar to that of
Ref.~\cite{d0ll425pbcsec}.
For the $e+$track channel, electrons are restricted to
$\left|\eta\right|<1.1$, and the leading jet must have $p_T>40$ GeV.
The dominant \ltrk\ background arises from $Z\to ee$ and $Z\to\mu\mu$
production, so we design the event selection to reject these events.

When the invariant mass of the lepton-track pair (\mlt) 
is in the range $70<$\mlt$<110$ GeV, the
\met\ requirement is tightened to \met\ $>35$ (40) GeV 
for the $e+$track ($\mu+$track) channel.
Furthermore, we introduce the variable \metzfit\ that corrects the \met\ in $Z\to\ell\ell$
events for mismeasured lepton momenta.  We rescale the lepton and track momenta 
according to their resolutions to bring \mlt\ 
to the mass of the $Z$ boson 
(91.2 GeV) and then use these rescaled momenta to correct the \met. 
Event selection based on this variable reduces the $Z$ background by half while providing
96\% efficiency for \ttbar\ events.
The cuts on \metzfit\ are always identical to those on \met.  

At least one jet is required to be identified as a $b$ jet
which provides strong background rejection for the \ltrk\ 
channels.  The $m_t$ precision is limited by signal statistics in the observed 
event sample when the background is reasonably low.
The above selection is a result of an optimization which minimizes the statistical uncertainty
on $m_t$. We do this in terms of \met, \metzfit, the transverse momenta of the leading
two jets, and the $b$-tagging criteria by stepping through two or more different
thresholds on these requirements. After considering all possible sets of selections, 
we choose the one which gives the best average expected statistical uncertainty on 
the $m_t$ measurement using many pseudoexperiments.
The expected statistical uncertainty varies smoothly over a 15\% range while
the study is sensitive to 5\% changes of the average statistical uncertainty.

We explicitly veto events satisfying the selection of any of the $2\ell$ channels, 
so the $\ell+$track channels are statistically independent of the $2\ell$ channels.
We select eight events in the $e+$track channel and six events in the
$\mu+$track channel.

{\subsection{Modeling Instrumental Backgrounds}
\label{sec:instrBG}}

Backgrounds can arise from instrumental effects in which the \met\ is mismeasured.
The main instrumental backgrounds for the $ee$, $\mu\mu$, $e+$track, and 
$\mu+$track channels are the $Z/\gamma^{*}\to ee$ and $Z/\gamma^{*}\to \mu\mu$ 
processes.  In these cases, apparent \met\ results from tails in jet or lepton $p_T$ 
resolutions. 
We use the NNLO cross section for $Z/\gamma^*\to ee,\mu\mu$ processes, along with
the Monte Carlo-derived efficiencies to estimate these backgrounds for
the $ee$ and $\mu\mu$ channels. The Monte Carlo kinematic distributions, including the \met, are verified
to reproduce a data sample dominated by these processes.
For the
\ltrk\ channels, we normalize Drell-Yan Monte Carlo so that the total expected event 
yield in a \ltrk\ sample with low \met\ equals the observed event yield in the data.
We observe a slightly different $p_T^Z$ distribution for simulated $Z\to
\ell\ell$ events in comparison with data.  As a result, all
$Z$ boson simulated samples, including the $Z\to\tau\tau$ physics
background samples,
are reweighted to the observed distribution of $p_T^Z$ in data~\cite{zpt}.

Another background arises when a lepton or a track within a jet 
is identified as an isolated lepton or track.  We utilize different
methods purely in data 
to estimate the level of these backgrounds for each channel.
In all cases, however, we distinguish reconstructed muons and tracks 
as ``loose" rather than ``tight" by releasing the isolation
criteria.  We make an analogous distinction for electrons by 
omitting the requirement on the electron
likelihood, $\mathcal L_e$, for ``loose'' electrons.

To determine the misidentified electron background yield in the $ee$ and $e\mu$ 
channels, we fit the observed distribution of $\mathcal L_e$ in the data to
a sum of the distributions from real isolated electrons and misidentified
electrons.  We determine the shape of $\mathcal L_e$ for
real electrons from a $Z\to ee$ sample with \met\ $<15$~GeV. 
For the $ee$ channel, we extract the shape for the misidentified 
electrons from a sample in which one ``tag electron" is required
to have both $\chi^2_{\rm \/ hmx}$ and $\mathcal L_e$ inconsistent with being from an 
electron.  We further require $M_{ee}<60$~GeV or $M_{ee}>130$~GeV  
and \met\ $<15$~GeV to reject $Z$~and $W$ boson events. 
The distribution of $\mathcal L_e$ is obtained from a separate ``probe electron"
in the same events.  In the $e\mu$ channel, the $\mathcal L_e$ distribution 
for misidentified electrons
is obtained in a sample with a non-isolated muon and \met\ $<15$~GeV.

To estimate the background from non-isolated muons for the $e\mu$ and $\mu\mu$ channels,
we use control samples to measure the fraction of muons, $f_{\mu}$, with 
$p_T>15$ GeV that appear to be isolated.
To enhance the heavy flavor content which gives non-isolated muons,
the control samples are selected to have two muons where a ``tag" muon is required to
be non-isolated.  We use another ``probe" muon to determine $f_{\mu}$.
The background yield for the $e\mu$ channel is computed from the number of events having 
an isolated electron, a muon with no isolation requirement, and the same sign charge for
the two leptons.  We multiply the observed yield by $f_{\mu}$.

We estimate the instrumental background for the $\mu\mu$ and \ltrk\ channels by
using systems of linear equations describing the composition of data samples with different
``loose" or ``tight" lepton and/or track selections.  We relate
event counts in these samples to the numbers of events with real or
misidentified isolated leptons using the system of equations.  These equations
take as inputs the efficiencies for real or misidentified leptons and tracks to pass the 
tight identification requirements.
For the $\mu\mu$ and \ltrk\ channels, we determine the efficiencies for 
real leptons and tracks to pass the tight
identification criteria using $Z\to ee$ and $Z\to\mu\mu$ events. 

For the \ltrk\ channels, the probabilities for misidentified leptons and tracks to pass the 
tight selection criteria are determined from 
multijet data samples with at least one loose lepton plus a jet.
We reject the event if two leptons of the same flavor satisfy tight criteria
to suppress
Drell-Yan events.  We also reject events with one or more tight leptons 
with different flavor from the loose lepton.  
These tight lepton vetos allow some events with two loose leptons or a lepton
and track in the sample.  We further suppress resonant $Z$ production by 
selecting events when \mlt\ and
$M_{\ell\ell} >$ 100 GeV or \mlt\ and $M_{\ell\ell} <$ 70 GeV.  
We reject $W$+jets events and misreconstructed $Z/\gamma^*$ events by
requiring \met\ $<$ 15 GeV and \metjes $<$ 25 GeV.  
Here, \metjes\ is the missing transverse energy with only jet energy corrections 
and no lepton corrections.  We use the latter because loose leptons no longer
adhere to standard resolutions.  We calculate the 
probability for electrons or muons to be misidentified by dividing
the number of tight leptons by the number of loose leptons.  For the track
probability, we combine the $e+$jet and $\mu+$jet samples and make the additional 
requirement that there be at least one loose track in the event.
The tight track misidentification probability is again the number of tight tracks
divided by the number of loose tracks.

To obtain samples dominated by misidentified isolated leptons
for mass analysis, we select events with two loose leptons or tracks plus
two jets.  For the $2\ell$ channels, we additionally require same sign dilepton
events.
\newline

\subsection{Composition of Selected Samples}

The expected numbers of background and signal events in all 
five channels (assuming a top quark production cross section of $7.0~\rm pb$)
are listed in Table~\ref{tab:evt_yields} along with 
the observed numbers of candidates. 
The $\mu+$track selection has half the efficiency of the $e+$track selection primarily due
to the tight $\mu\mu$ veto.  The expected and observed event yields agree for all channels.

\begin{table*}[htp]
\caption{\label{tab:evt_yields} Expected event yields for \ttbar\ (we assume $\sigma_{t\bar{t}} = 7.0\;\rm pb$) 
and backgrounds and numbers of 
observed events for the five channels. The $2\ell$ channel uncertainties include statistical as well 
as systematical uncertainties while the $e+$track and $\mu+$track uncertainties are statistical only.}
\begin{tabular}{lrcclrclcrcl rcl rclc}
\hline \hline
 Sample & \multicolumn{3}{c}{$t\overline{t}$} & \quad\quad &\multicolumn{3}{c}{Diboson}    & \quad\quad & \multicolumn{3}{c}{$Z$} & \multicolumn{3}{c}{Multijet/$W+$jets} & \multicolumn{3}{c}{Total} & Observed \\
\hline
 $e\mu$ & $36.7 $&$\pm$&$ 2.4 $  &  &$1.7 $&$\pm$&$ 0.7$  && $4.5 $&$\pm$&$ 0.7$ & $\quad\quad2.6$&$\pm$&$ 0.6$ & $44.5$&$\pm$&$2.7$ & 39 \\
 $ee$ & $11.5 $&$\pm$&$ 1.4$  & &$0.5$&$\pm$&$ 0.2$ && $2.3 $&$\pm$&$ 0.4$ & $\quad\quad0.6 $&$\pm$&$ 0.2$ &  $14.8$&$\pm$&$1.5$ & 17 \\
 $\mu\mu$ & $8.3 $&$\pm$&$ 0.5 $& &$0.7$&$\pm$&$ 0.1 $ && $4.5$&$\pm$&$ 0.4 $ & $\quad\quad0.2$&$\pm$&$ 0.2 $& $13.7$&$\pm$&$0.7 $ & 13 \\
 $e+$track & $9.4$&$\pm$&$0.1$ && $0.1$&$\pm$&$ 0.0$ &&$0.4$&$\pm$&$ 0.1$ & $\quad\quad0.4$&$\pm$&$0.1$&  $10.3$&$\pm$&$0.2$ & 8 \\
 $\mu+$track & $4.6$&$\pm$&$0.1$ && $0.1$&$\pm$&$0.0$ && $0.7$&$\pm$&$0.1$ & $0.1$&$\pm$&$0.0$& $5.5$&$\pm$&$0.1$   & 6 \\
\hline \hline
\end{tabular}
\end{table*}

\begin{figure*}
\includegraphics{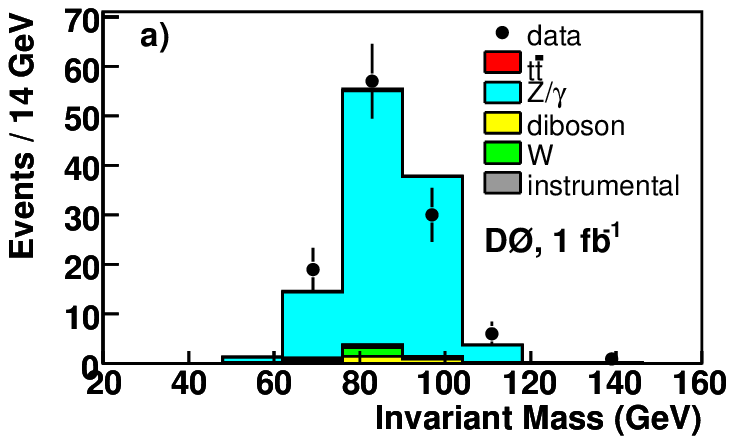}
\includegraphics{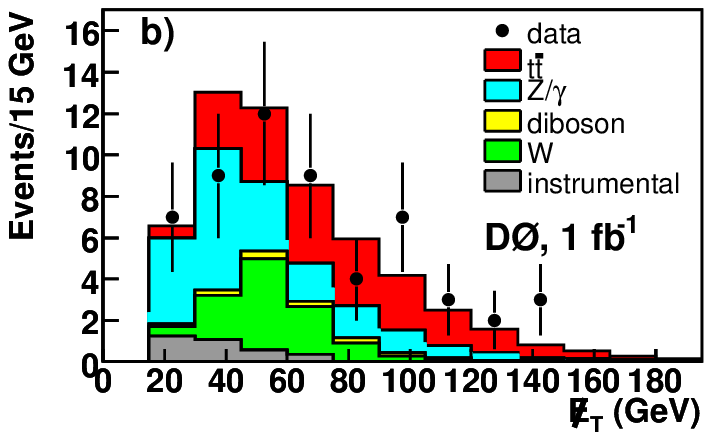}
\caption{\label{fig:LepTrack} Comparison of the expected distributions
from backgrounds and \ttbar\ ($m_t=170$ GeV) in the \ltrk\ channels. (a) \mlt\ for 
the $e+$track channel without the requirement of the
$b$-tag and with inverted \met\ cuts.  (b) \met\ for the sum of both \ltrk\ channels,
again without the $b$-tag requirement. We assume $\sigma_{t\bar{t}} = 7.0\;\rm pb$.}
\end{figure*}

Kinematic comparisons between data and the sum of the signal and background expectations
provide checks of the content and properties of our data sample. 
Figure~\ref{fig:LepTrack}(a) shows the expected and observed distributions
of \mlt\ in the $e+$track channel without the $b$-tag requirement and for
an inverted \met\ requirement. The $\mu+$track distribution looks similar (not shown). 
The mass peak at $M_Z$ indicates the $e+$track sample is primarily composed
of $Z\to ee$ events before the final event selection.  In Fig.~\ref{fig:LepTrack}(b), we
show the \met\ distribution in the $\ell+$track channels
after all cuts except the $b$-tag requirement.

\begin{figure*}[!ht]
\includegraphics{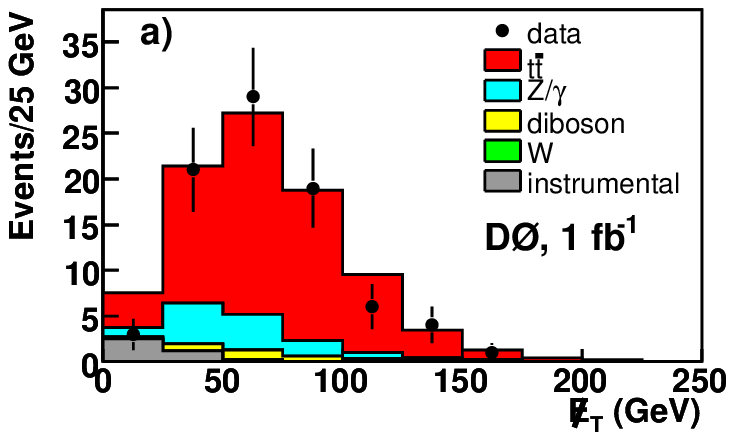}
\includegraphics{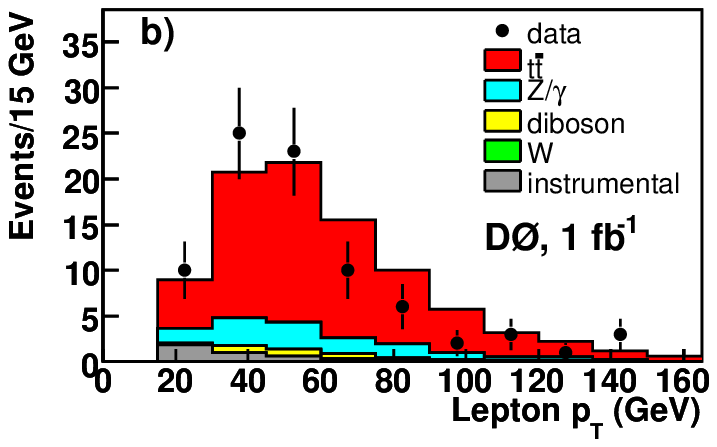}
\caption{\label{fig:PtsAllCuts} (a) \met\  and (b) leading lepton $p_T$ for 
\ttbar\ ($m_t=170$ GeV) and background processes
overlaid with those for observed events in all channels after final event selection. 
We assume $\sigma_{t\bar{t}} = 7.0\;\rm pb$.}
\end{figure*}

The expected numbers of background and signal events after all selections in all
five channels are listed in Table~\ref{tab:evt_yields} along with
the observed numbers of candidates.  We assume $\sigma(\ttbar)=7.0$ pb.
We do not include systematic uncertainties for the $\ell+$track channels.
The small backgrounds mean their uncertainties have a negligible effect on the measured
$m_t$ uncertainty.
The $\mu+$track selection has half the efficiency of the $e+$track selection primarily due
to the tight $\mu\mu$ veto.  The expected and observed event yields agree for all channels.
Figures~\ref{fig:PtsAllCuts}(a) and (b) show the
\met\ and leading lepton $p_T$ summed over all 
channels for the final candidate sample.  
We observe the data distributions to agree with our signal and background model.

{\section{Event Reconstruction}
\label{sec:kinRec}}

Measurement of the dilepton event kinematics and constraints from the $t\bar{t}$
decay assumption allow a partial reconstruction of the
final state and a determination of $m_t$.
Given the decay of each top quark to a $W$ boson and a $b$ quark, with each 
$W$ boson decaying to a charged lepton and a neutrino, there are six final state 
particles: two
charged leptons, two neutrinos, and two $b$ quarks.  Each particle
can be described by three momentum components.  Of these eighteen independent  
parameters, we can directly measure only the momenta of the leptons.  The leading 
two jets most often come from the $b$ quarks.  Despite final state radiation and 
fragmentation, 
the jet momenta are highly correlated with those of the underlying $b$ quarks.  
We also measure 
the $x$ and $y$ components of the \met, \mex\ and \mey, from the neutrinos.  This leaves
four quantities unknown.  We can supply two constraints 
by relating the four-momenta of the leptons and neutrinos to the masses
of the $W$ bosons:
\begin{linenomath}
\begin{eqnarray}
\begin{gathered}
M_W^{2}=(E_{\nu_1}+E_{l_1})^{2}-(\vec{p}_{\nu_1}+\vec{p}_{l_1})^{2} \\
M_W^{2}=(E_{\nu_2}+E_{l_2})^{2}-(\vec{p}_{\nu_2}+\vec{p}_{l_2})^{2},
\end{gathered}
\label{eq:wmasses}
\end{eqnarray}
\end{linenomath}
\noindent  where the subscript indices indicate the $\ell\nu$ pair coming from one
or another $W$ boson.
Another constraint is supplied by requiring that the mass of the
top quark and the mass of the anti-top quark be equal:
\begin{linenomath}
\begin{equation}
\begin{gathered}
(E_{\nu_1}+E_{l_1}+E_{b_1})^{2}-(\vec{p}_{\nu_1}+\vec{p}_{l_1}+\vec{p}_{b_1})^{2}=\\
(E_{\nu_2}+E_{l_2}+E_{b_2})^{2}-(\vec{p}_{\nu_2}+\vec{p}_{l_2}+\vec{p}_{b_2})^{2}.
\label{eq:teqtbar}
\end{gathered}
\end{equation}
\end{linenomath}

\noindent  The last missing constraint can be supplied by a hypothesized value
of the top quark mass.  With that, we can solve the equations and calculate
the unmeasured top quark and neutrino momenta that are consistent with the
observed event.  Usually, the
dilepton events are kinematically consistent with a large range of $m_t$.  
We quantify this consistency, or ``weight," for each $m_t$ by testing
measured quantities of the event (e.g., \met\ or lepton and jet $p_T$) against
expectations from the dynamics of \ttbar\ production and decay. 
This requires us to sample from relevant \ttbar\ distributions, yielding
many solutions for a specific $m_t$.
We sum the weights for each solution for each $m_t$.
The distribution of weight vs.~$m_t$ is termed a ``weight distribution" of a given event.
Using parameters from these weight distributions,
we can then determine the most likely value of $m_t$.

Several previous efforts to measure $m_t$ using dilepton events have used
event reconstruction techniques.  The differences between methods stem largely
from which event parameters are used to calculate the event weight.
We use the \nuwt\ and MWT  
techniques to determine the weighting as described below.
\newline

\subsection{Neutrino Weighting}

The  \nuwt\ method omits the measured \met\ for kinematic reconstruction.  
Instead, we choose the pseudorapidities of the two neutrinos from \ttbar\ decay 
from their expected distributions.  
We obtain the distribution of neutrino $\eta$ from several simulated \ttbar\
samples with a range of $m_t$ values.  These distributions
can each be approximated by a single 
Gaussian function.   The standard deviation specifying this function varies weakly
with $m_t$.
Once the neutrino pseudorapidities are fixed and a value for $m_t$ assumed, we can solve for the complete
decay kinematics, including the unknown neutrino momenta.  There may be
up to four different combinations of solved neutrino momenta
for each assumed pair of neutrino $\eta$ values for each event.  
We assume the leading two jets are the $b$ jets, so there are two
possible associations of $W$ bosons with $b$ jets.  

For each pairing of neutrino momentum solutions, we define a weight, $w$,
based on the agreement between the measured
\met\ and the sum of the neutrino momentum components in $x$ and $y$, $p_x^{\nu}$ and 
$p_y^{\nu}$.
We assume independent Gaussian resolutions in measuring \mex\ and \mey.
The weight is calculated as
\begin{linenomath}
\begin{equation}\label{eq:nuweight}
w=\exp\left[\frac{-(\not\!\!E_{x}^{\rm \/ calc}-\not\!\!E_{x}^{\rm obs})^2}{2(\sigma^{\rm \/ u}_{x})^2}\right]\exp\left[\frac{-(\not\!\!E_{y}^{\rm \/ calc}-\not\!\!E_{y}^{\rm \/ obs})^2}{2(\sigma^{\rm \/ u}_y)^2}\right],
\end{equation}
\end{linenomath}
reflecting the agreement between the measured and calculated \met.
\meiobs\ ($i=x$ or $y$) are the components of the measured event \met, and \meicalc\ are 
the components of the \met\
calculated from the neutrino transverse momenta resulting from each
solution.  We calculate the quantities \meiunc\ to be the 
sums of the energies projected onto the $i$ axes measured by all ``unclustered'' 
calorimeter cells -- those cells not included in jets or electrons.
The high $p_T$ objects, leptons and jets, enter into the determination of
both \meicalc\ and \meiobs\ whereas the unclustered energy \meiunc\ only enters
into \meiobs.  Given the resolutions $\sigma^{\rm \/ u}_i$ of the \meiunc, we can 
therefore estimate the probability that the \meiobs\ are consistent with the
\meicalc\ from the \ttbar\ hypothesis.

\begin{figure}
\includegraphics{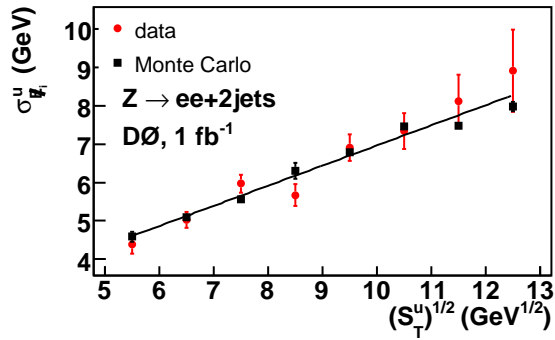}
\caption{\label{fig:metset}Dependence of the resolution of unclustered \met\ on 
the unclustered scalar transverse missing energy for $Z\to ee$ events with exactly two jets.}
\end{figure}

As parameters of the method, we determine $\sigma^{\rm \/ u}_i$ using 
$Z\rightarrow ee+2$~jets data and Monte Carlo events.  
We calculate an unclustered scalar transverse energy, $S_T^{\rm u}$, as the scalar 
sum of the transverse energies of all unclustered calorimeter cells.
Due to the azimuthal isotropy of the calorimeter, we
observe that the independent $x$ and $y$ components of the $\sigma^{\rm \/ u}_i$
depend on $S_T^{\rm u}$ in the same way within their uncertainties.
Therefore, we combine results for both components to determine our resolution 
more precisely.
We find agreement between data and simulation in the observed dependence of
these parameters on $S_T^{\rm u}$.  
The distributions are shown for these combined resolutions in Fig. \ref{fig:metset}.
We fit the unclustered \met\ resolutions obtained from simulation as
\begin{linenomath}
\begin{equation}
\sigma_x^{\rm \/ u}(S_T^{\rm \/ u})=\sigma_y^{\rm \/ u}(S_T^{\rm \/ u})=4.38 {\rm~GeV} +0.52 \sqrt{S_T^{\rm \/ u}} {\rm~GeV},
\end{equation}
\end{linenomath}
\noindent  and use this parametrization for 
the unclustered missing energy resolution for both data and Monte Carlo in 
Eq.~(\ref{eq:nuweight}).

For each event, we consider ten different $\eta$ assumptions for
each of the two neutrinos.  We extract these values from the
histograms appropriate to the $m_t$ being assumed.
The ten $\eta$ values are the medians of each of ten ranges of
$\eta$ which each represent 10\% of the \ttbar\ sample for
a given $m_t$.

\subsection{Matrix Weighting}

In the MWT approach, we use the measured momenta of the two 
charged leptons. We assign the measured momenta of the two jets with the 
highest transverse momenta to the $b$ and $\bar{b}$ quarks and the measured \met\ 
to the sum of the transverse momenta of the two neutrinos from the decay of the 
$t$ and $\bar{t}$ quarks. We then assume a top quark mass and determine the 
momenta of the $t$ and $\bar{t}$ quarks that are consistent with these 
measurements. We refer to each such pair of momenta as a solution for the 
event. For each event, there can be up to four solutions.

We assign a weight to each solution, analogous to the \nuwt\ weight of Eq.~\ref{eq:nuweight},
given by
\begin{linenomath}
\begin{equation}
\label{eq:mxweight}
w=f(x)f(\bar{x})p(E_{\ell}^*|m_t)p(E_{\bar{\ell}}^*|m_t),
\end{equation}
\end{linenomath}
where $f(x)$ is the PDF for the proton for 
the momentum fraction $x$ carried by the initial quark, and $f(\bar{x})$ is 
the corresponding value for the initial antiquark. The quantity $E_{\ell}^*$ is the observed
lepton energy in the top quark rest frame.
We use the central fit of the \cteq\ PDFs and evaluate them at $Q^2=m_t^2$. 
The quantity $p(E_{\ell}^*|m_t)$ in Eq.~\ref{eq:mxweight} is the probability that for the hypothesized top 
quark mass $m_t$, the lepton $\ell$ has the measured $E_{\ell}^*$ \cite{dalitz}:
\begin{linenomath}
\begin{equation}
p(E_{\ell}^*|m_t)=\frac{4m_tE_{\ell}^*(m_t^2-m_b^2-2m_tE_{\ell}^*)}{(m_t^2-m_b^2)^2+M_W^2(m_t^2-m_b^2)-2M_W^4}.
\label{eq.elProb}
\end{equation}
\end{linenomath}

\subsection{Total Weight vs. $m_t$}

Equations~\ref{eq:nuweight} and \ref{eq:mxweight} indicate how the event weight
is calculated for a given top quark mass in the \nuwt\ and MWT methods.  In each
method, we consider all solutions and jet assignments to get a total weight, 
$w_{\rm tot}$, for a given $m_t$.  In general, 
there are two ways to assign the two jets to the $b$ and $\bar{b}$ quarks.  There are
up to four solutions for each hypothesized value of the top 
quark mass.  The likelihood for each assumed 
top quark mass $m_t$ is then given 
by the sum of the weights over all the possible solutions:
\begin{linenomath}
\begin{equation}
w_{\rm \/ tot} = \sum_i \sum_j w_{ij},
\end{equation}
\end{linenomath}
\noindent where $j$ sums over the solutions for each jet assignment $i$.  
We repeat this calculation for both the \nuwt\ and MWT methods
for a range of assumed top quark masses
from 80 GeV through 330 GeV.

For each method, we also account for the finite resolution of jet and lepton 
momentum measurements.  We repeat the weight 
calculation with input values for the 
measured momenta (or inverse momenta for muons) drawn from normal distributions centered 
on the measured
values with widths equal to the known detector resolutions.
We then average the weight distributions obtained from $N$ such variations:
\begin{linenomath}
\begin{equation}
w_{\rm \/ tot}(m_t)=N^{-1}\sum_{n=1}^N w_{{\rm \/ tot},n}(m_t).
\end{equation}
\end{linenomath}
where $N$ is the number of samples.
One important benefit of this procedure is that the efficiency
of signal events to provide solutions increases.  For instance, the \nuwt\ 
efficiency to find a solution for \ttbar $\rightarrow e\mu$ events is 95.9\% 
without resolution sampling, 
while 99.5\% provide solutions when $N=150$.  
For the MWT analysis, events with 
$m_t=175$ GeV yield an efficiency of 90\% without resolution sampling.  This
rises to over 99\% when $N=500$.  We use $N=150$ and 500 for \nuwt\ and
MWT, respectively.

\begin{figure}
\includegraphics{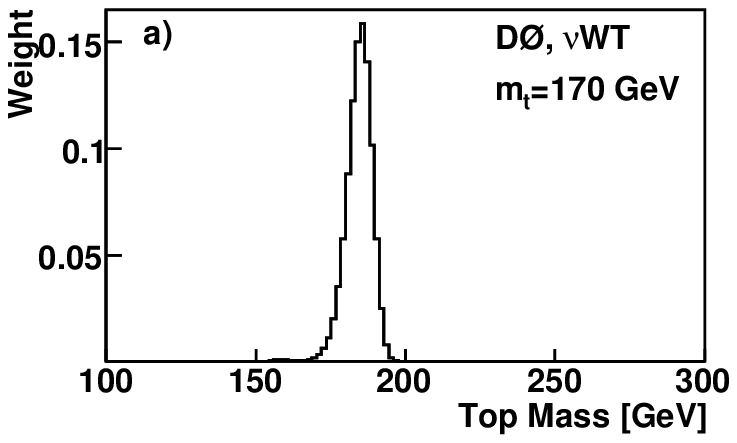}
\includegraphics{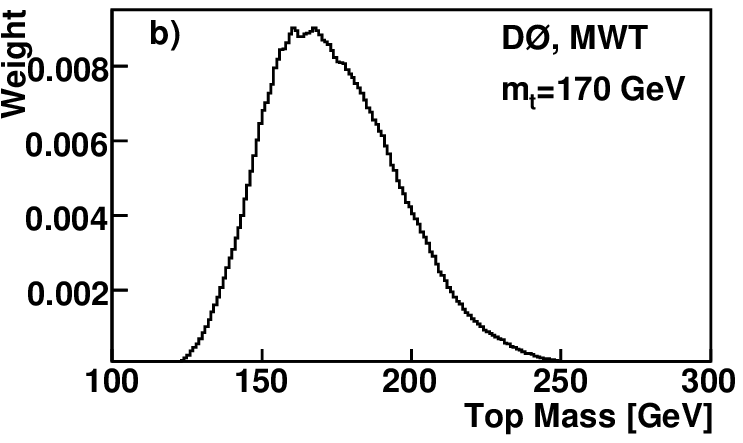}
\caption{\label{fig:weight} Example weight distributions for different
single $t\bar{t}\rightarrow e\mu$
Monte Carlo events obtained with a) \nuwt\ method and b) MWT method. The generator
level mass is $m_t=170$ GeV.}
\end{figure}

Examples of a single event weight distribution for \nuwt\ and MWT
are shown in Fig. ~\ref{fig:weight} for two different simulated events.
The sensitivities of the two methods are
similar on average, but the different widths of the weight distributions vary
significantly in both approaches on an event-by-event basis.  This can be caused
by an overall
insensitivity of an event's kinematic quantities to $m_t$, or to a different
sensitivity when using those kinematic quantities with specific
event reconstruction techniques.

\begin{figure*}
\includegraphics[scale=1.0]{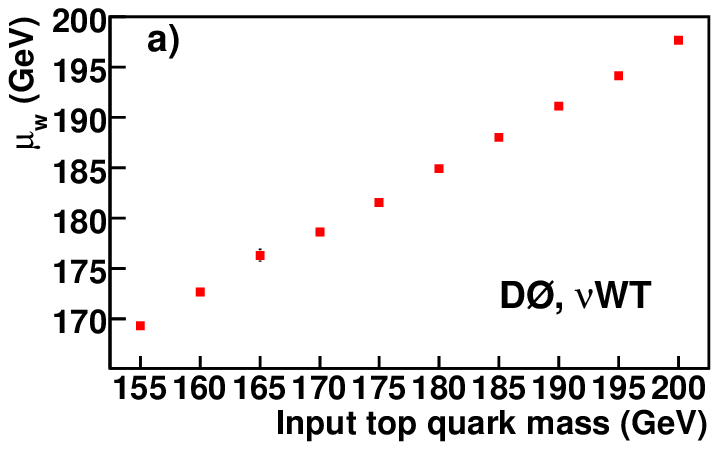}
\includegraphics[scale=1.0]{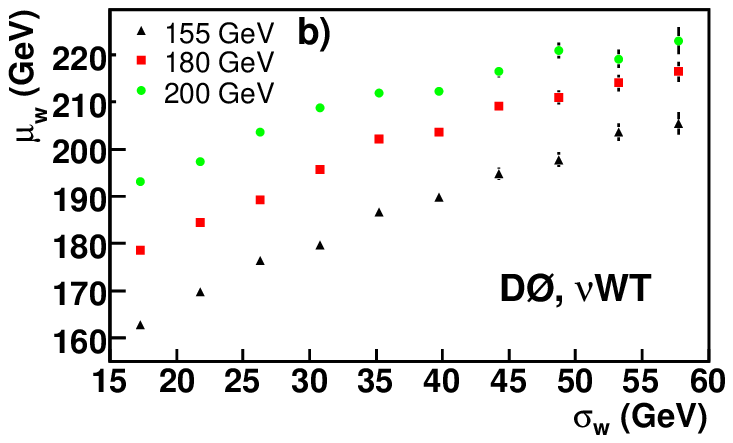}
\caption{\label{fig:moments} (a) Correlation
between the mean of the \nuwt\
weight distribution and the input $m_t$.  (b) Correlation between
\nuwt\ $\mu_w$ and $\sigma_w$ for the $e\mu$ channel.
Three test masses of 155 GeV, 180 GeV, and 200 GeV are shown.}
\end{figure*}

Properties of the weight distribution are strongly correlated with 
$m_t$ if the top quark decay is as expected in the standard model.  For instance, 
Fig.~\ref{fig:moments}(a) illustrates the correlation of the mean of the 
\nuwt\ weight distribution, $\mu_w$, with the generated top 
quark mass from the Monte Carlo.  The relationship between the root-mean-square
of the weight distribution, $\sigma_w$, and $\mu_w$ also varies with $m_t$, as
shown in Fig.~\ref{fig:moments}(b).
There is the potential for non-standard decays of the top quark.  
For $m_t=170$~GeV and assuming BR$(H^{\pm}\to \tau \nu) \sim 100\%$,
we observe $\mu_w$ (\nuwt) to shift systematically upward
when a $H^{\pm}$ boson of mass 80~GeV is present in the decay 
chain instead of a $W$ boson.  
When BR$(t\to H^{\pm}b)=$100\%, this shift is 10\%. Thus, the measurements of this paper 
are strictly valid only for standard model top quark decays.

{\section{Extracting the Top Quark Mass}
\label{sec:LikeHood}}

We cannot determine the top quark mass directly from $\mu_w$ or from
the most probable mass from the event weight distributions, max$_w$.  
Effects such as initial and final state radiation systematically shift 
these quantities from the actual top quark mass. 
In addition, the presence of background must be taken into
account when evaluating events in the candidate
sample.  We therefore perform a maximum likelihood fit to distributions
(``templates") of characteristic variables from the weight distributions.
This fit accounts for the shapes of signal templates and for the 
presence of background.
\newline

\subsection{Measurement Using Templates}
\label{sec:pdhlike}
\begin{figure*}[htp]
\includegraphics{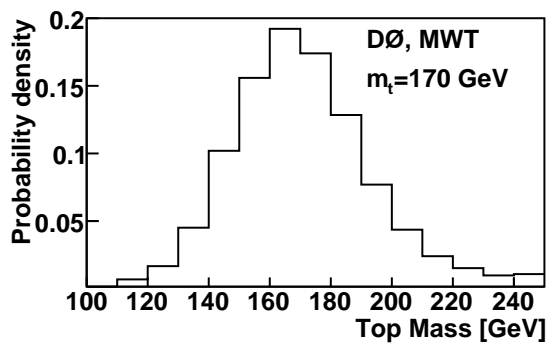}
\caption{\label{fig:MWTtemplate} An example of a template for the MWT method.}
\end{figure*}

We define a set of input variables
characterizing the weight distribution for event $i$, denoted by $\{x_i\}_N$, 
where $N$ is the number of variables.
Examples of $\{x_i\}_N$ might be the integrated weight in bins of a coarsely binned 
template, or they might
be the moments of the weight distribution.
A probability density histogram for simulated signal 
events, $h_s$, is defined as an $(N+1)$-dimensional 
histogram of input top quark mass vs.~$N$ variables.  For background, $h_b$
is defined as an $N$-dimensional histogram of the $\{x_i\}_N$.
Both $h_s$ and $h_b$ are normalized to unity:
\begin{linenomath}
\begin{eqnarray}
\int h_s (\{x_i\}_N \mid m_t)\, d \{x_i\}_N = 1, \\
\int h_b (\{x_i\}_N)\,d\{x_i\}_N = 1.
\end{eqnarray}
\end{linenomath}
An example of a template for the MWT method is shown on
Fig.~\ref{fig:MWTtemplate}.
We measure $m_t$ from $h_s (\{x_i\}_N,m_t)$ and 
$h_b(\{x_i\}_N)$ using a maximum likelihood method.
For each event in a given data sample, 
all $\{x_i\}_N$ are found and used for the likelihood calculation.
We define a likelihood $\mathcal L$ as
\begin{linenomath}
\begin{eqnarray}
\begin{gathered}
\mathcal L(\{x_i\}_N,n_b,N_{\rm obs} \mid m_t) = \\
\prod_{i=1}^{N_{\rm \/ obs}}\frac{n_s h_s(\{x_i\}_N \mid
m_t)+n_b h_b(\{x_i\}_N)}{n_s+n_b},
\label{eq:Lhisto}
\end{gathered}
\end{eqnarray}
\end{linenomath}
where $N_{\rm obs}$ is the number of events in the sample, $n_b$ is the number
of background events, and $n_s$ is the signal event yield. 
We obtain a histogram of $-\ln{\mathcal L}$ vs.~$m_t$ for the sample.  We 
fit a parabola that is symmetric around the point with the highest likelihood 
(lowest $-\ln{\mathcal L}$).  The fitted mass range is
several times larger than the expected statistical uncertainty. 
It is chosen $a$ $priori$ to give the best sensitivity to the top quark mass using Monte Carlo
pseudoexperiments, and is typically around $\pm 20$ GeV.

We obtain measurements of $m_t$ for several channels by multiplying
the likelihoods of these channels:
\begin{linenomath}
\begin{equation}
-\ln{\mathcal L} = \sum_{\rm \/ ch} ( -\ln{\mathcal L}_{\rm \/ ch} ) \;\;,
\label{eq:lhChans}
\end{equation}
\end{linenomath}
where ``ch" denotes the set of channels.  In this paper, we calculate overall likelihoods
for the $2\ell$ subset, $\rm \/ ch $~$\in \lbrace e\mu,ee,\mu\mu\rbrace$;
the $\ell+$track subset, $\rm \/ ch$~$ \in \lbrace e+$track, $\mu+$track$\rbrace$; and
the five channel dilepton set, $\rm \/ ch$~$ \in \lbrace e\mu,ee,\mu\mu,e+$track, $\mu+$track$\rbrace$.

\subsection{Choice of Template Variables}
The choice of variables 
characterizing the weight distributions has
been given some consideration in the past.  
For example, the \dzero\ MWT analysis and CDF \nuwt\ analyses
have used max$_{w}$~\cite{r2CDFnuWT,r2d0mt}.  
Earlier \dzero\ \nuwt\ analyses
employed a multiparameter probability density technique 
using the coarsely binned weight distribution
to extract a measure
of $m_t$~\cite{r1d0mt,r2d0mt}

For the MWT analysis described here, we use the single parameter approach.
In particular, to extract the mass, we use Eq.~\ref{eq:Lhisto} where 
$x_i = \lbrace$max$_{w} \rbrace$.  We determine
the values of $n_s$ and $n_b$ by scaling the sum of the 
expected numbers of signal ($\bar{n}_s$) and background ($\bar{n}_b$)
events in Table~\ref{tab:evt_yields}
to the number observed in each channel.  We fit the histogram
of $-\ln{\mathcal L}$ to a parabola using a 40 GeV
wide mass range centered at the top quark mass 
with the minimal value of $-\ln{\mathcal L}$ for all channels.

For the \nuwt\ analysis, we define the optimal set of input 
variables for the top quark mass
extraction to be the set that simultaneously minimizes the expected statistical 
uncertainty and the number of variables.
The coarsely binned weight distribution approach exhibits up to 20\% better
statistical performance than single parameter methods for a given kinematic reconstruction
approach by using more information.  
However, over five bins are typically needed, and the large number of variables
and their correlations significantly complicate the analysis.
We study the performance of the method with many different choices of variables from
the weight distributions.  These include single parameter choices such as 
max$_{w}$ or $\mu_w$, which provide similar performance 
in the range 140 GeV $<m_t< 200$ GeV.

Vectors of
multiple parameters included various coarsely binned templates, or subsets
of their bins.  For the \nuwt\ analysis, we observe 
that individual event weight distributions
have fluctuations which are reduced by considering bulk properties
such as their moments.
The most efficient parameters are the
first two moments ($\mu_w$ and $\sigma_w$) of the weight 
distribution.  This gives 16\%
smaller expected statistical uncertainty than using max$_{w}$ or $\mu_w$ alone.
The improvement of the performance comes from the fact that $\sigma_w$ is 
correlated with $\mu_w$ for a given input top quark mass.
This is shown in Fig.~\ref{fig:moments}(b) for three different input top quark masses.
The value of $\sigma_w$ helps to better identify the range of input $m_t$ that is
most consistent with the given event having a specific $\mu_w$.  This ability to 
de-weight incorrect $m_t$ assignments results in a narrower likelihood 
distribution and causes a corresponding reduction in the statistical uncertainty.
No other choice of variables gives significantly better performance.
The use of the weight distribution moments, $x_i = \lbrace \mu_w, \sigma_w \rbrace$, 
in $h_s$ and $h_b$ with Eq. \ref{eq:Lhisto} is termed \nuwth.

Because the templates are two-dimensional for background and three-dimensional for signal, 
a small number of bins are unpopulated.  
We employ a constant extrapolation for unpopulated edge bins using 
the value of the populated 
bin closest in $m_t$ but having the same $\mu_w$ and $\sigma_w$.
For empty bins flanked by populated bins in the $m_t$ direction but
with the same $\mu_w$ and $\sigma_w$, we employ a linear interpolation.
We fit the histogram of $-\ln{\mathcal L}$ vs. $m_t$ with a 
parabola.  When performing the fit, the \nuwth\ approach determines $n_s$ to be 
$n_s=N_{\rm obs}-\bar{n}_b$.  Fit ranges of 50, 40, and 30 GeV are used for \ltrk, $2\ell$, 
and all dilepton channels, respectively.

\subsection{Probability Density Functions}
\begin{figure*}
\includegraphics[width=150mm,height=75mm]{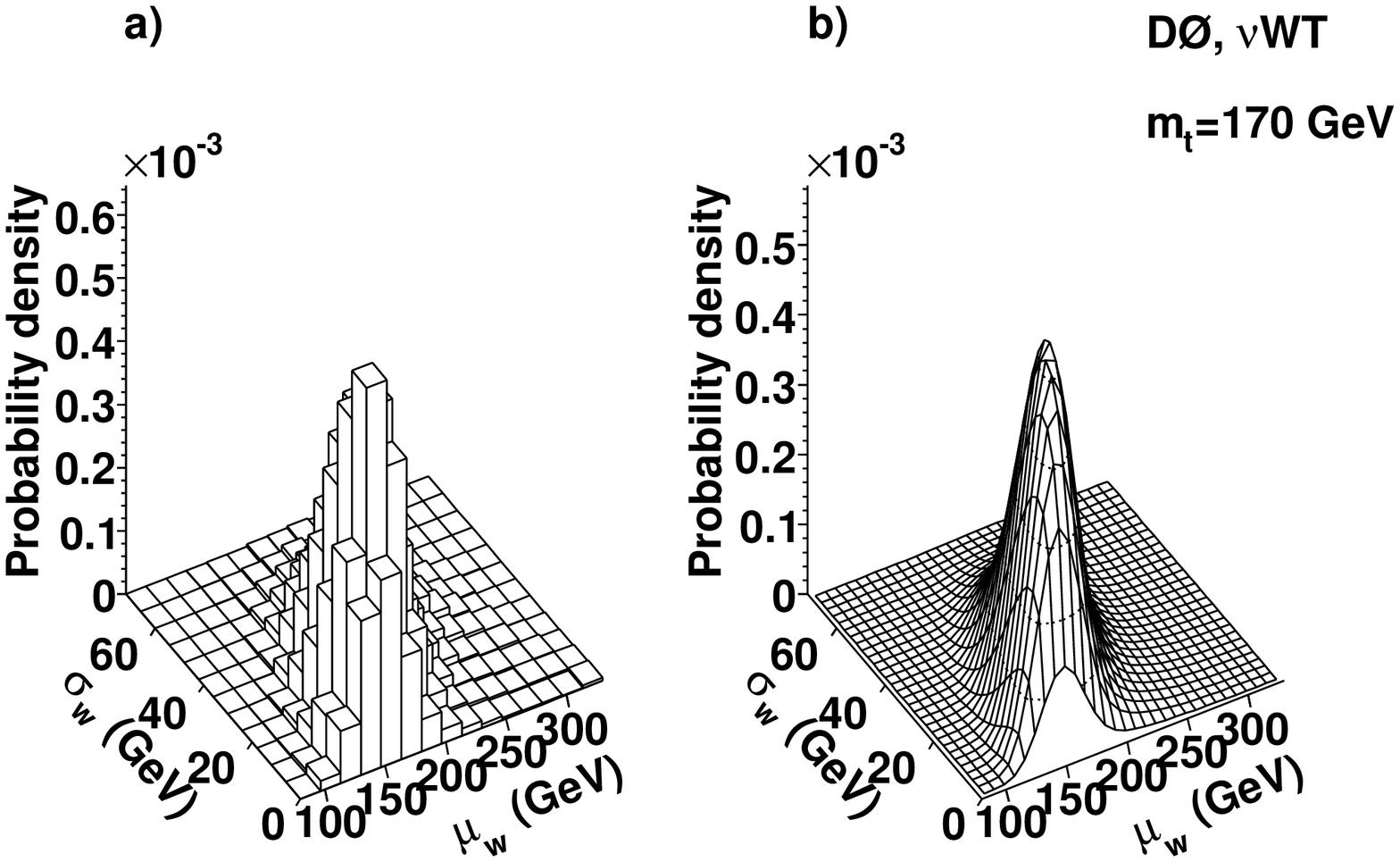}
\includegraphics[width=150mm,height=75mm]{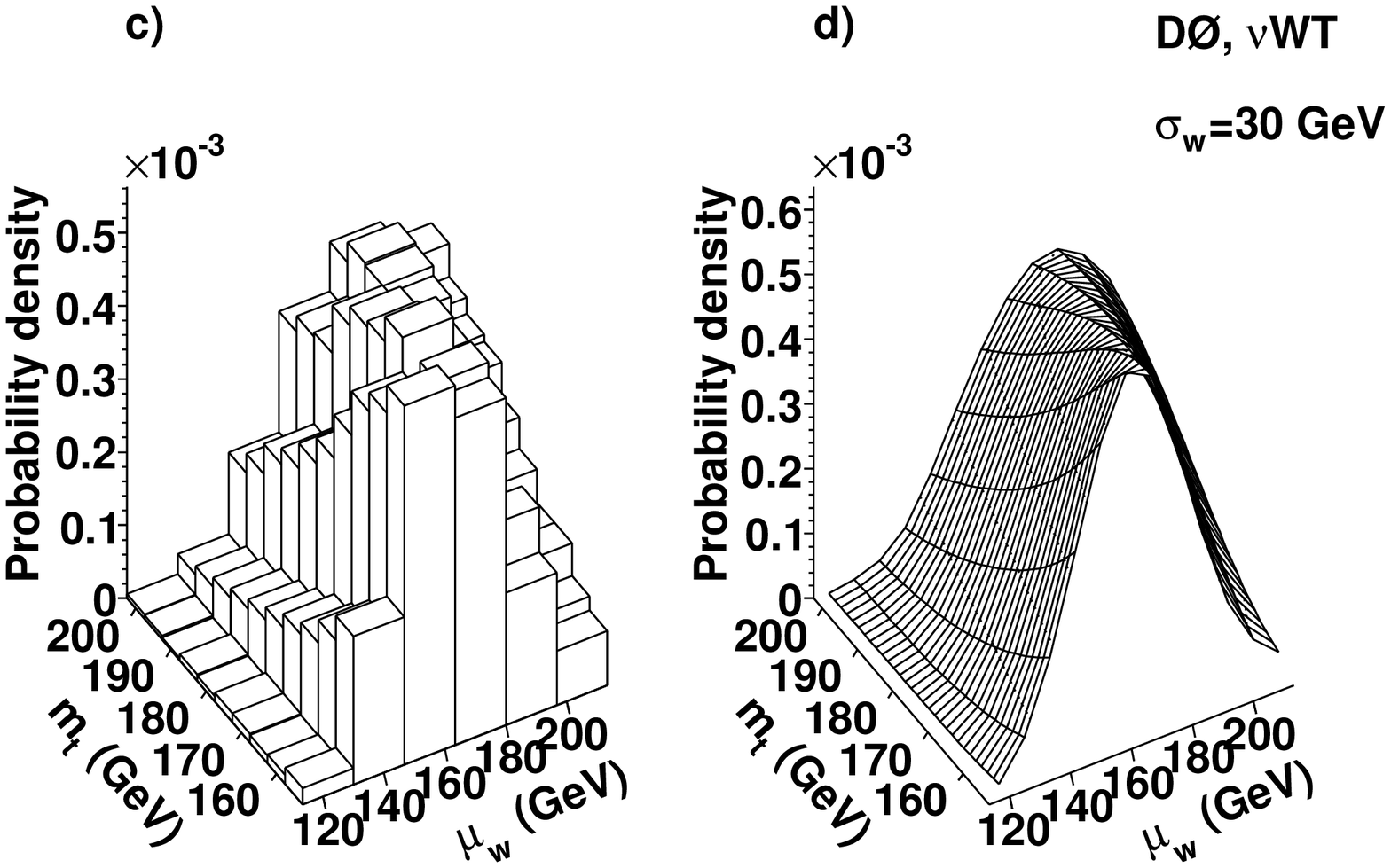}
\caption{\label{fig:slice}  
Slices of probability density histograms $h_s$ and fit functions $f_s$ for 
the \nuwt\ method in the $e\mu$ channel.  Probability densities vs.
$\sigma_w$ and $\mu_w$ for $m_t=170$~GeV are shown for (a) $h_s$ and (b) $f_s$.
Probability densities vs. $m_t$ and $\mu_w$ for $\sigma_w=30$~GeV 
are shown for (c) $h_s$ and (d) $f_s$.}
\end{figure*}

In both methods described above, 
there are finite statistics in the simulated samples used to model $h_s$ and $h_b$, 
leading to bin-by-bin fluctuations. We address this in the \nuwt\ analysis 
by performing fits to $h_s$ and $h_b$ templates.
We term this version of the \nuwt\ method \nuwtf.  
For the signal, we generate a probability density function $f_s$ by fitting $h_s$ with 
the functional form
\begin{linenomath}
\begin{equation}
\begin{gathered}
\label{eq:pdf}
f_s(\mu_w,\sigma_w \mid m_t)= (\sigma_w+p_{13})^{p_6} \exp[-p_7(\sigma_w+p_{13})^{p_8}] \\
\times \biggl\{ (1-p_{9}) \frac{1}{\sigma\sqrt{2\pi}} \exp[-\frac{(\mu_w-m)^2}{2\sigma^2}]\\
+\,p_{9}\cdot\frac{p_{11}^{1+p_{12}}}{\Gamma(1+p_{12})}(\mu_w-\frac{m}{p_{10}})^{p_{12}}\exp[-p_{11}(\mu_w - \frac{m}{p_{10}})] \\
\times\,\Theta(\mu_w - \frac{m}{p_{10}}) \biggr\} \\
\times \biggl\{\int_{0}^{\infty}  
(x+p_{13})^{p_6}\exp[-p_7(x+p_{13})^{p_8}] dx \biggr\}^{-1}.
\end{gathered}
\end{equation}
\end{linenomath}
The parameters $m$ and $\sigma$ are linear functions of $\sigma_w$ and $m_t$: 
\begin{linenomath}
\begin{equation}
\begin{gathered}
\label{eq:MandS}
 m=p_0+p_1(\sigma_w -36~{\rm GeV}) +p_2 (m_t-170~{\rm GeV}), \\
\sigma=p_3+p_4(\sigma_w -36~{\rm GeV}) +p_5 (m_t-170~{\rm GeV}). \\
\end{gathered}
\end{equation}
\end{linenomath}
Equations~\ref{eq:pdf} and ~\ref{eq:MandS} are $ad$ $hoc$ functions 
determined empirically.  A typical $\chi^2$ with respect to $h_s$ is
found in the $e\mu$ channel which yields 4.0 per degree of freedom.
The linear relationship between $\sigma_w$ and $\mu_w$ is shown in 
Fig.~\ref{fig:slice}(a), which is an example of the probability density
vs. $\sigma_w$ and $\mu_w$ for fixed input top quark mass of 170 GeV.
The dependence of $f_s$ on $\sigma_w$ is expressed in the first line 
of Eq.~\ref{eq:pdf}. The second and third lines contain a Gaussian plus an 
asymmetrical function to describe the dependence on $\mu_w$. The
factors $1/(\sigma\sqrt{2\pi})$ and $p_{11}^{1+p_{12}}/\Gamma(1+p_{12})$ 
and the integral in the fourth line normalize the probability density 
function to unity.  Examples of 
two-dimensional slices of three-dimensional signal histograms for fixed 
input $m_t = 170$~GeV and $\sigma_w=30$~GeV are shown in 
Fig.~\ref{fig:slice}(a) and (c), respectively.  The corresponding
slices of the fit functions are shown in Fig.~\ref{fig:slice}(b) and (d),
respectively.

\begin{figure*}
\includegraphics[width=150mm,height=75mm]{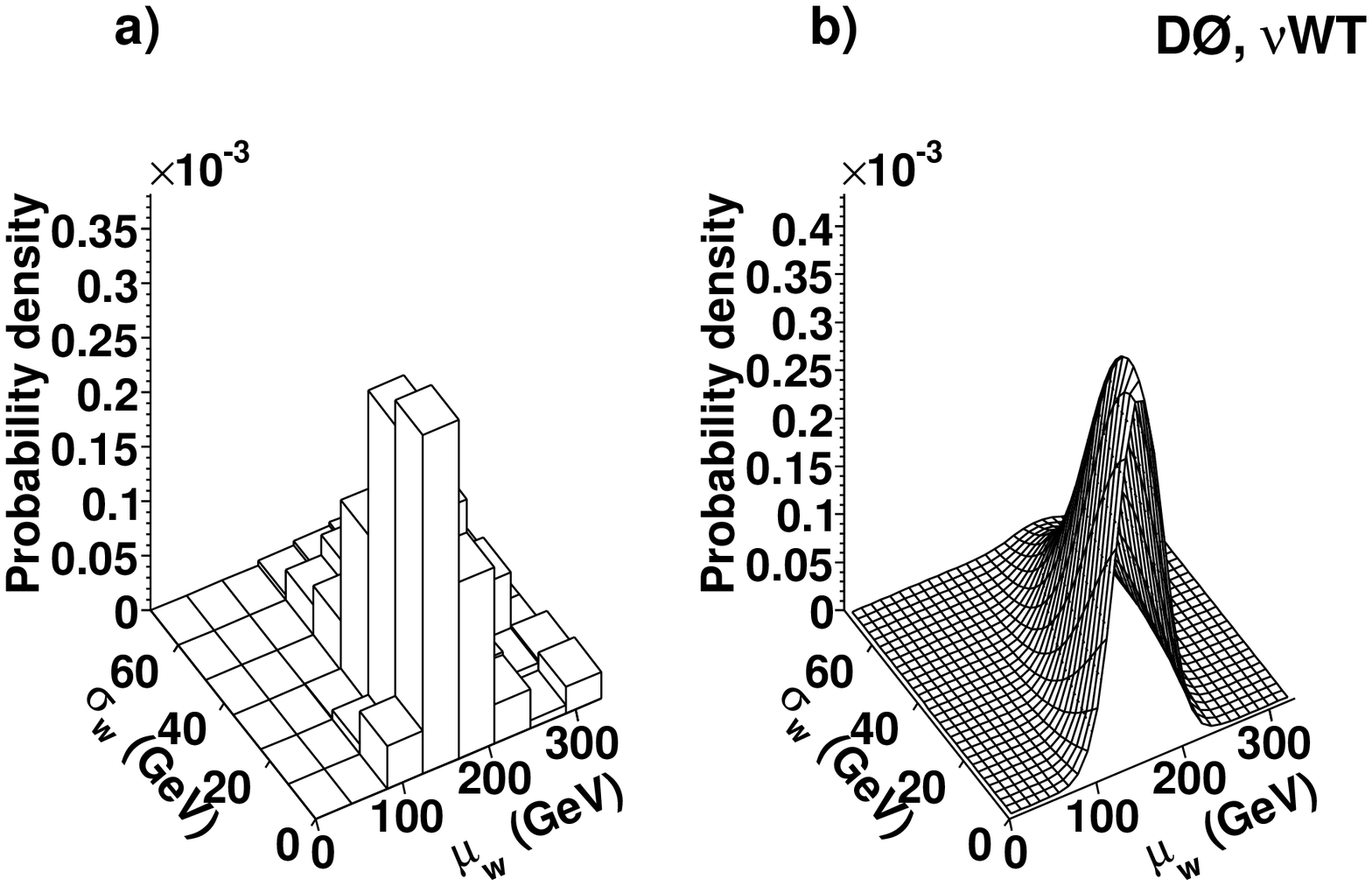}
\caption{\label{fig:bgslice}
Probability density histogram $h_b$ and fit function $f_s$ for
the \nuwt\ method in the $e\mu$ channel.  Probability densities vs.
$\sigma_w$ and $\mu_w$ are shown for (a) $h_b$ and (b) $f_b$.}
\end{figure*}

The background probability density function $f_b(\mu_w,\sigma_w)$ is obtained as
the normalized two-dimensional function of $\mu_w$ and $\sigma_w$ of simulated background events: 
\begin{linenomath}
\begin{equation} 
\begin{gathered}
f_b(\mu_w, \sigma_w) = \\
\frac{ \exp\left[-(p_1\mu_w +p_2\sigma_w-p_0)^2-(p_4\mu_w + p_5 \sigma_w -p_3)^2\right]}
{\int_0^{\infty}\int_0^{\infty}\exp\left[-(p_1 x +p_2 y-p_0)^2  -(p_4 x + p_5 y -p_3)^2\right]d\it{x}\,d\it{y}}.
 \end{gathered}
\end{equation}
\end{linenomath}
This is also an $ad$ $hoc$ function determined empirically.  The fit is performed
to $h_b$ containing the sums of all backgrounds for each channel and according
to their expected yields.
A typical $\chi^2$ with respect to $h_b$ is found in the $e\mu$ case which yields
5.2 per degree of freedom.  Examples of $h_b$ and $f_b$ are shown in
Fig.~\ref{fig:bgslice}(a) and (b), respectively.

To measure $m_t$, we begin by adding two extra terms to the likelihood $\mathcal L$ of Eq.~\ref{eq:Lhisto}.
The first term is a constraint that requires that the fitted sum of the number of signal events $n_s$ and the 
number of background events $n_b$ agrees within Poisson fluctuations with the number of observed events $N_{\rm \/ obs}$:
\begin{linenomath}
\begin{equation}
  \mathcal L_{\rm poisson}(n_s+n_b,N_{\rm \/ obs})\equiv\frac{(n_s+n_b)^{N_{\rm \/ obs}}\exp[-(n_s+n_b)]}{N_{\rm \/ obs}!}.
\end{equation}
\end{linenomath}
The second term is a Gaussian constraint that requires agreement between the 
fitted number of background events $n_b$ and the number of expected background 
events $\bar{n}_b$ within Gaussian fluctuations, where the width of the Gaussian 
is given by the estimated uncertainty $\delta_b$ on $\bar{n}_b$: 
\begin{linenomath}
\begin{equation}
  \mathcal L_{\rm gauss}(n_b,\bar{n}_b,\delta_b)\equiv\frac{1}{\sqrt{2\pi}\delta_b}\exp{[-(n_b-\bar{n}_b)^2/2\delta_b^2]}.
\end{equation}
\end{linenomath}
The total likelihood for an individual channel is given by
\begin{linenomath}
\begin{eqnarray}
\begin{gathered}
\mathcal L({\mu_w}_i,{\sigma_w}_i,\bar{n}_b,N_{\rm \/ obs} \mid m_t,n_s,n_b) = \\
\mathcal L_{\rm gauss}(n_b,\bar{n}_b,\delta_b)\mathcal L_{\rm poisson}(n_s+n_b,N_{\rm \/ obs})\\
\times\prod_{i=1}^{N_{\rm \/ obs}}\frac{n_sf_s({\mu_w}_i, {\sigma_w}_i \mid
m_t)+n_bf_b({\mu_w}_i, {\sigma_w}_i)}{n_s+n_b}.
\label{eq:Lfit}
\end{gathered}
\end{eqnarray}
\end{linenomath}
The product extends over all events in the data sample. The maximum of the 
likelihood corresponds to the measured top quark mass. We simultaneously minimize 
$-\ln{\mathcal L}$ with respect to  $m_t$, $n_s$, and $n_b$ 
using \minuit~\cite{root,minuit}. The fitted sample composition is consistent with the expected one.
 This yields an analytic function of 
$-\ln{\mathcal L}$ vs. $m_t$.
Its statistical uncertainty is found by fixing $n_s$ 
and $n_b$ to their optimal values and taking half of distance between the points at 
which the $-\ln\mathcal L$ value is 0.5 units greater than its minimum value. 
We obtain measurements of $m_t$ for several channels 
by minimizing the combined 
$-\ln{\mathcal L}$ simultaneously with respect to $m_t$ and the numbers of signal and 
background events for the channels considered.

\subsection{Pseudoexperiments and Calibration}

The maximum likelihood fits attempt to account for the presence of
background and for the signal and background shapes of the templates.
For a precise measurement of $m_t$, we must test for any residual
effects that can cause a shift in the relationship between the fitted and
actual top quark masses.  We test our fits and extract correction factors 
for any observed shifts by performing pseudoexperiments.

A pseudoexperiment for each channel 
is a set of simulated events of the same size and composition 
as the selected dataset given in Table \ref{tab:evt_yields}.
We compose it by randomly drawing simulated events out of the 
large Monte Carlo event pool.
Within a given pool, each Monte Carlo event has a weight based on production
information and detector performance parameters.  An example of
the latter is the
$b$-tagging efficiency which depends on jet $p_T$ and $\eta$, and an 
example of former is the weight with which each event is generated by \alpgen.
We choose a random event, 
and then accept or 
reject it by comparing the event weight to a random number.  
In this way, our pseudoexperiments are
constructed with the mix of events that gives the correct kinematic 
distributions.

For MWT, we compose pseudoexperiments by
drawing Monte Carlo events from signal and background samples with
probabilities proportional to the numbers of events expected, $\bar{n}_s$ and
$\bar{n}_b$.  Thus, we draw events for each source based on
a binomial probability.  In the \nuwt\ pseudoexperiments, 
the number of background events of each source is Poisson 
fluctuated around the expected yields of Table~\ref{tab:evt_yields}. 
The remaining events in the pseudoexperiment are signal events. If the sum of backgrounds
totals more than $N_{\rm \/ obs}$, then the extra events are dropped and
$n_s = 0$.
In this way, we do not use the $t\bar{t}$ production cross section, 
which is a function of $m_t$. 

\begin{figure*}[tb!]
 \begin{center}
 \centering
  \includegraphics[height=55mm,width=0.3\textwidth]
                  {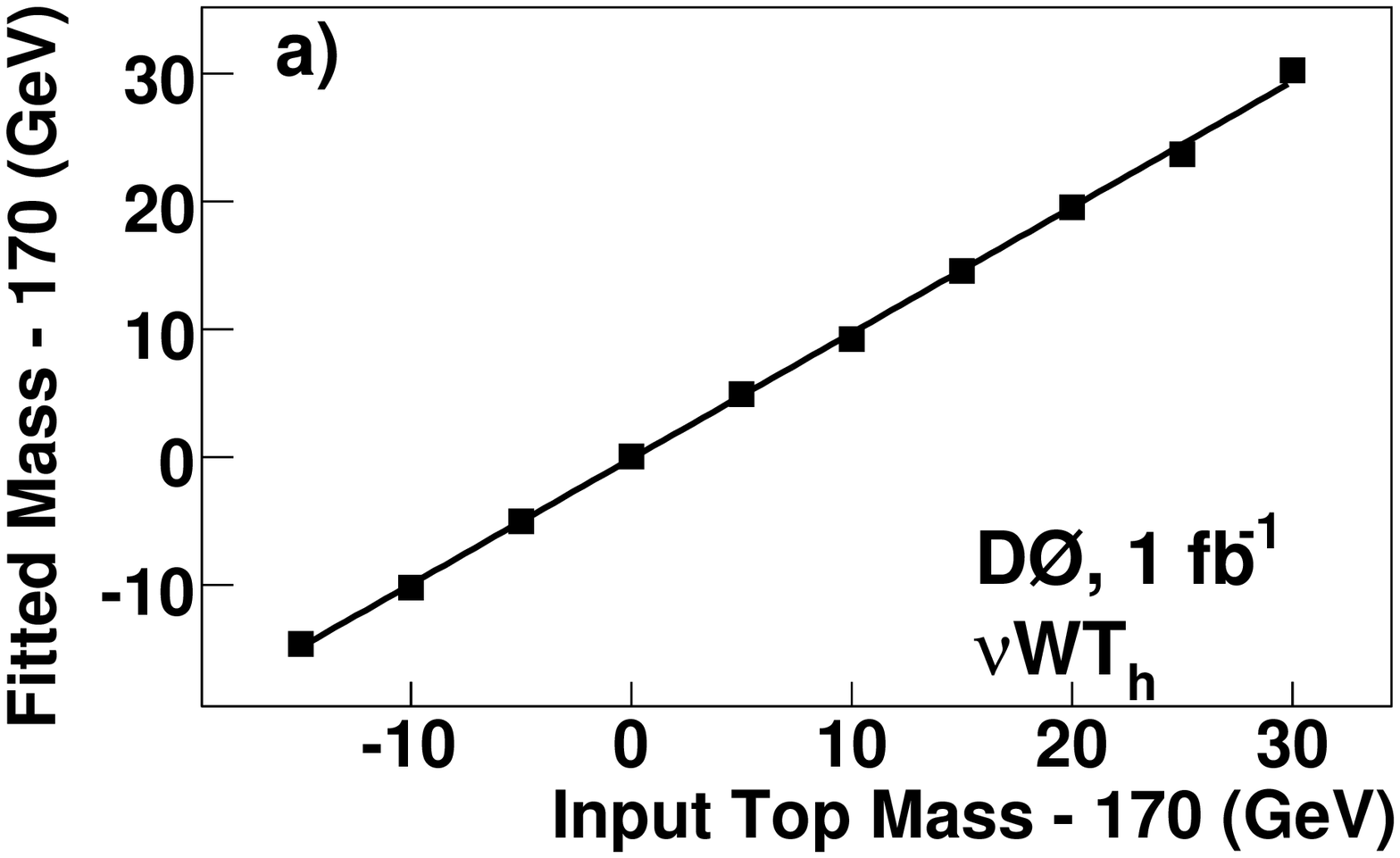} \hfill
  \includegraphics[height=55mm,width=0.3\textwidth]
                  {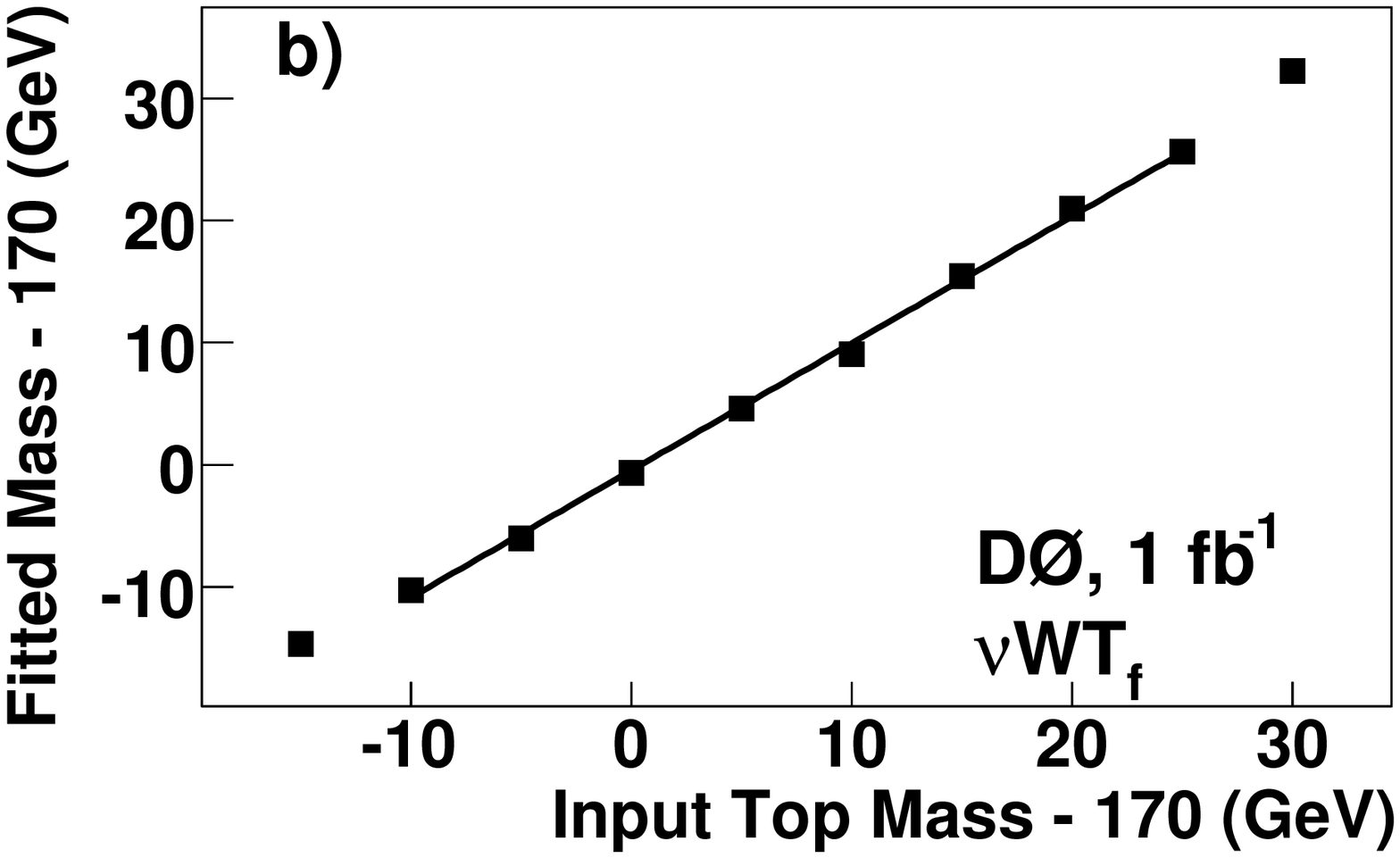} \hfill
  \includegraphics[height=55mm,width=0.3\textwidth]
                  {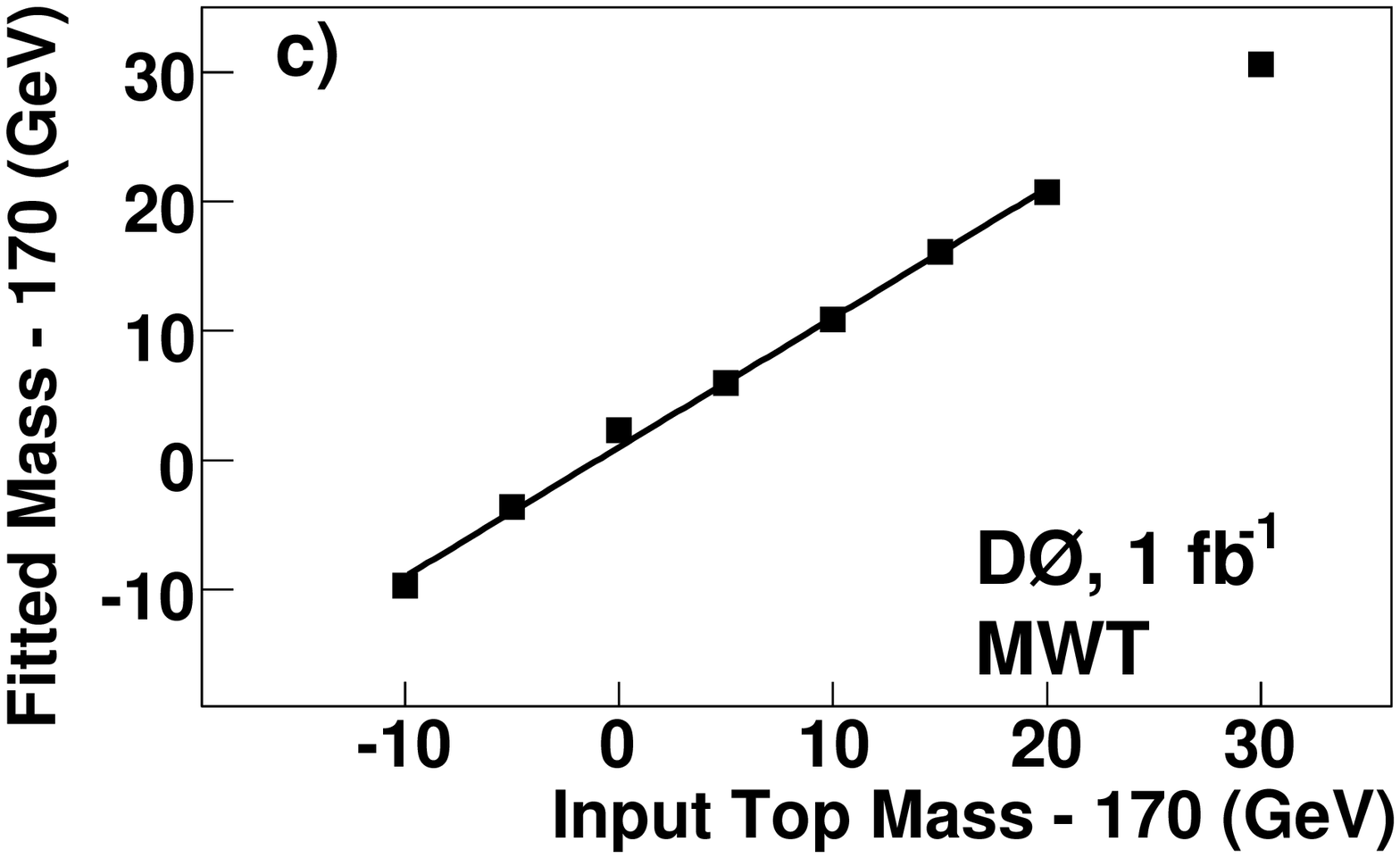} \hfill
  \hspace*{0.49\textwidth}
  \caption{The combined calibration curves corresponding to the (a) \nuwth, (b) \nuwtf, and (c) MWT methods. 
 Overlaid is the result of the linear fit as defined in Eq.~\ref{eqn:calibration}. The uncertainties are 
small and corresponding bars are hidden by the markers.}
  \label{fig:calibration_combined}
 \end{center}
\end{figure*}

To establish the relationship between the fitted top quark mass, \mtfit,
and the actual generated
top quark mass $m_t$, we assemble a set of many pseudoexperiments 
for each input mass.  For the \nuwt\ method, we choose 300 pseudoexperiments
because that is the point at which the Monte Carlo 
statistics begin to be oversampled,
and all pseudoexperiments can still be considered statistically independent.  
We average
\mtfit\ for each input $m_t$ for each channel.  We combine channels
according to Eq.~\ref{eq:lhChans}, and we fit
the dependence of this average mass on $m_t$ with
\begin{linenomath}
\begin{equation}
\langle m_t^{\rm fit}\rangle = \alpha (m_t- 170\; {\rm GeV}) + 
                   \beta  + 170  \; {\rm GeV}.
\label{eqn:calibration}
\end{equation}
\end{linenomath}
\noindent  The calibration points and fit functions are shown in 
Fig.~\ref{fig:calibration_combined}.
The results of the fits are summarized in Table~\ref{tab:calibration}.  
Ideally, $\alpha$ and $\beta$ should be unity
and zero, respectively.  The \mtfit\ of each pseudoexperiment and data measurement is 
corrected for the slopes and offsets given in Table~\ref{tab:calibration} by
\begin{linenomath}
\begin{equation}
m_t^{\rm meas} = \alpha^{-1} (m_t^{\rm fit}- \beta  - 170\; {\rm GeV})
                 + 170  \; {\rm GeV}.
\label{eqn:calibration2}
\end{equation}
\end{linenomath}

\begin{table*}
\begin{center}

\caption{Slope ($\alpha$) and offset ($\beta$)
from the linear fit in Eq.~\ref{eqn:calibration} to the pseudoexperiment results of 
Fig.~\ref{fig:calibration_combined} for the $2\ell$, \ltrk, and combined dilepton channel sets 
using the MWT and \nuwt\ methods.}
\begin{tabular}{ccccccccccccr@{.}ll}
\hline\hline
Method& Channel & \multicolumn{3}{c}{Slope: $\alpha$} & \multicolumn{3}{c}{Offset: $\beta$ [GeV]} & \multicolumn{3}{c}{Pull width} & \multicolumn{3}{c}{Expected statistical}\\
&	&		  &			  &            & & & & & & & \multicolumn{3}{c}{uncertainty [GeV]}\\
\hline
\nuwth&$2\ell$      & 0.98 &$\pm$& 0.01\phantom{0}  & $\;-$0.04 &$\pm$& 0.11 & $\quad$1.02 &$\pm$& 0.02 &$\quad\quad$ &5&8 	\\
\nuwth&\ltrk\       & 0.92 &$\pm$& 0.02\phantom{0}  & $\quad$2.28 &$\pm$& 0.27  & $\quad$1.04 &$\pm$& 0.02 & & 13&0	\\
\nuwth&combined     & 0.99 &$\pm$& 0.01\phantom{0}  & $\;-$0.04 &$\pm$& 0.11 & $\quad$1.03 &$\pm$& 0.02 & &5&1	\\
\hline
\nuwtf&$2\ell$      & 1.03 &$\pm$& 0.01\phantom{0}  & $\;-$0.32 &$\pm$& 0.15 & $\quad$1.06 &$\pm$& 0.02 & &5&8 	\\
\nuwtf&\ltrk\       & 1.07 &$\pm$& 0.03\phantom{0}  & $\;-$0.04 &$\pm$& 0.37 & $\quad$1.07 &$\pm$& 0.02 & &12&9	\\
\nuwtf&combined     & 1.04 &$\pm$& 0.01\phantom{0}  & $\;-$0.45 &$\pm$& 0.13 & $\quad$1.06 &$\pm$& 0.02 & &5&3	\\
\hline
MWT&$2\ell$         & 1.00 &$\pm$& 0.01 \phantom{0} & $\quad$0.95 &$\pm$& 0.05  &$\quad$ 0.98 &$\pm$& 0.01 & &6&3 	\\
MWT&\ltrk\          & 0.99 &$\pm$& 0.01 \phantom{0} & $\quad$0.64 &$\pm$& 0.12  &$\quad$ 1.06 &$\pm$& 0.01 & &13&8	\\
MWT&combined       & 0.99 &$\pm$& 0.01 \phantom{0} & $\quad$0.97 &$\pm$& 0.05  &$\quad$ 0.99 &$\pm$& 0.01 & &5&8	\\
\hline
\hline\hline
\end{tabular}
\label{tab:calibration}
\end{center}
\end{table*}

\begin{figure*}[htp]
 \begin{center}
 \centering
  \includegraphics[scale=0.35]{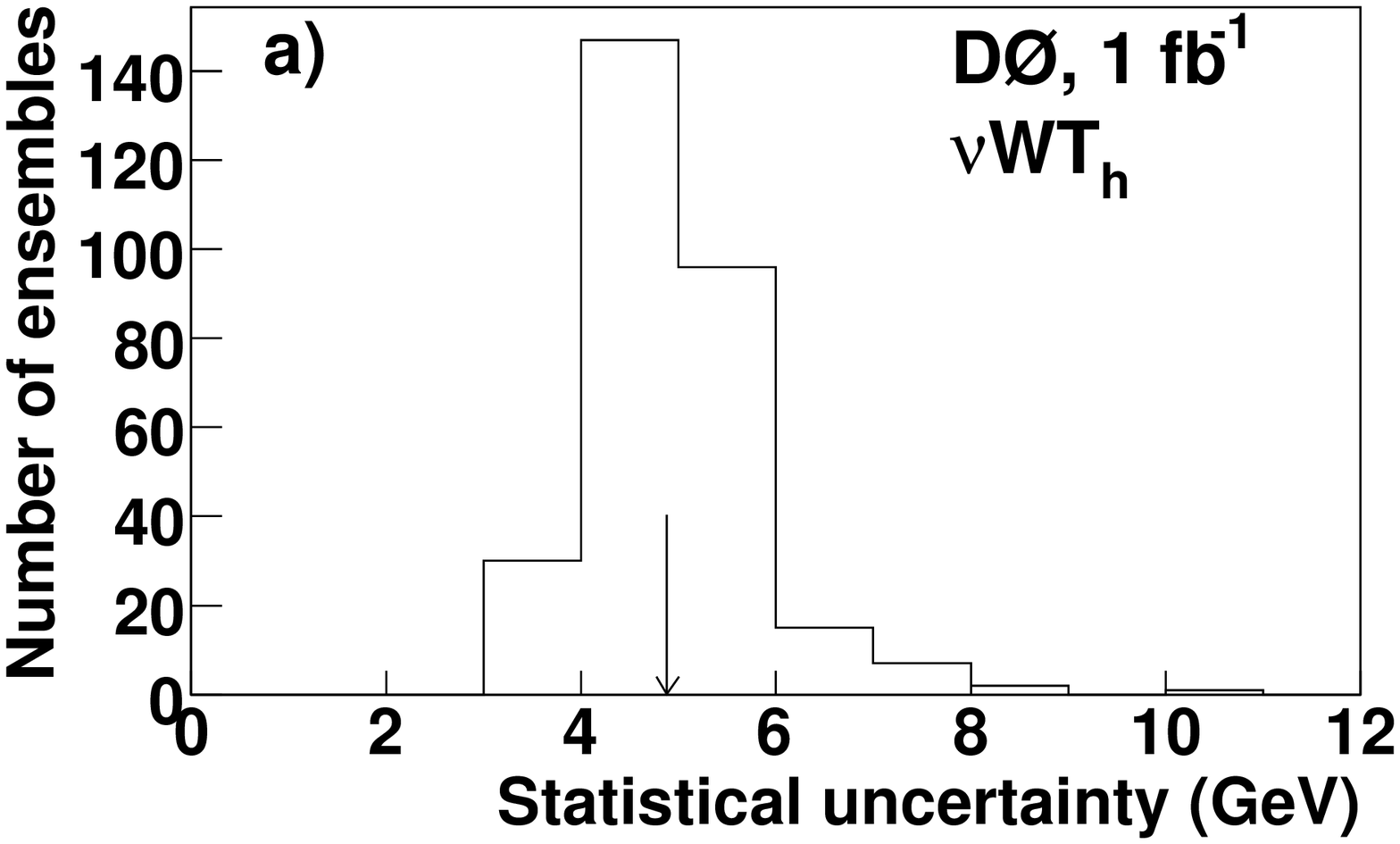}
  \includegraphics[scale=0.35]{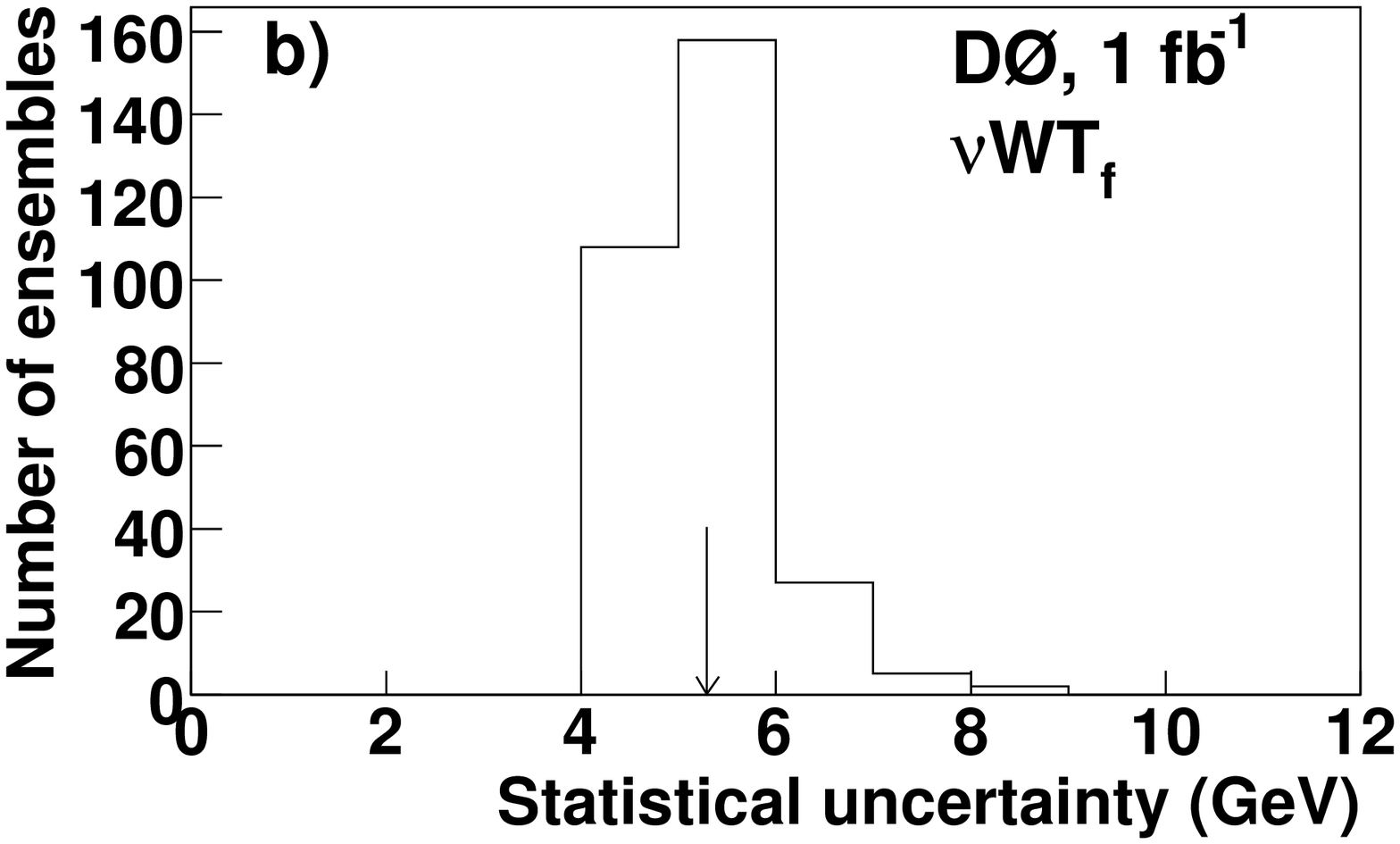}
  \includegraphics[scale=0.35]{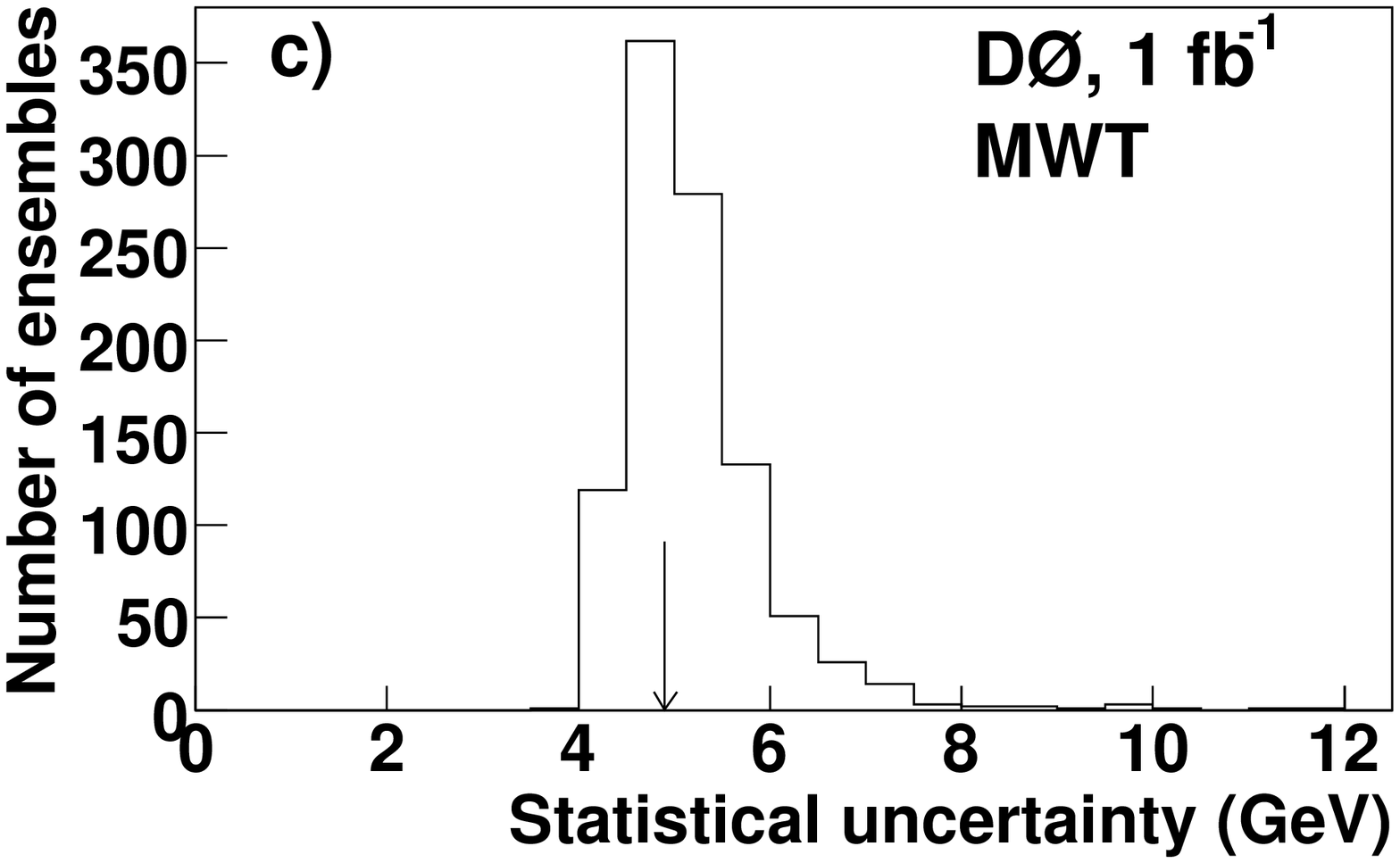}
    \caption{Distribution of statistical uncertainties for top quark mass measurements of 
pseudoexperiments for the combination of all channels for simulated events 
with $m_t=170~\rm GeV$ for the (a) \nuwth, (b) \nuwtf, and (c) MWT methods.
The uncertainties are corrected by the calibration curve and for the pull width. 
The arrows indicate the statistical uncertainties for the measured top quark mass.}
\label{fig:staterr_combined}
 \end{center}
\end{figure*}

The pull is defined as
\begin{linenomath}
\begin{equation}
{\rm pull} = \frac{m_t^{\rm meas}-m_t}{\sigma(m_t^{\rm meas})}, 
\end{equation}
\end{linenomath}
where $\sigma(m_t^{\rm meas})=\alpha^{-1} \sigma(m_t^{\rm fit})$ is the measured statistical
uncertainty after the calibration of Eq.~\ref{eqn:calibration2}. 
The ideal pull distribution has a Gaussian shape with a mean of zero and a 
width of one.  The pull widths from pseudoexperiments are given in Table~\ref{tab:calibration}.
A pull width larger (less) than one indicates an underestimated (overestimated) 
statistical uncertainty. The uncertainty of the data measurement is corrected 
for deviations of the pull width from one as well as for the slope of the calibration curve.
The mean of the distribution of calibrated and pull-corrected statistical 
uncertainties yields the expected statistical uncertainty (see Table~\ref{tab:calibration}). 
Figure~\ref{fig:staterr_combined} shows the pull-width-corrected distribution 
of statistical uncertainties for the $m_t$ 
measurements from the ensemble testing. 
The expected uncertainty on the combined measurement for all channels is 5.1, 5.3, and
5.8 GeV for \nuwth, \nuwtf, and MWT, respectively.

\begin{figure*}
\includegraphics[scale=0.5]{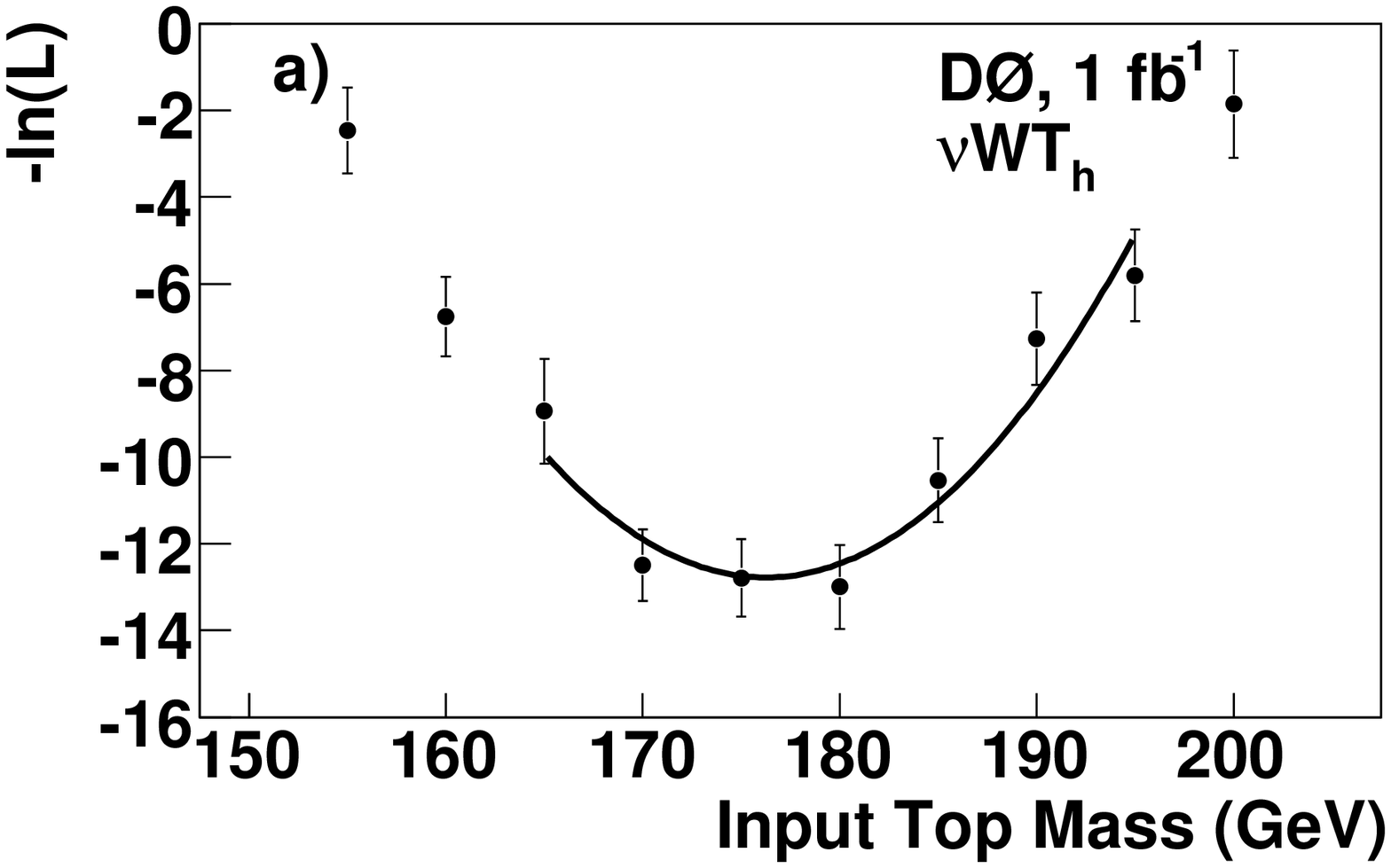}
\includegraphics[scale=0.5]{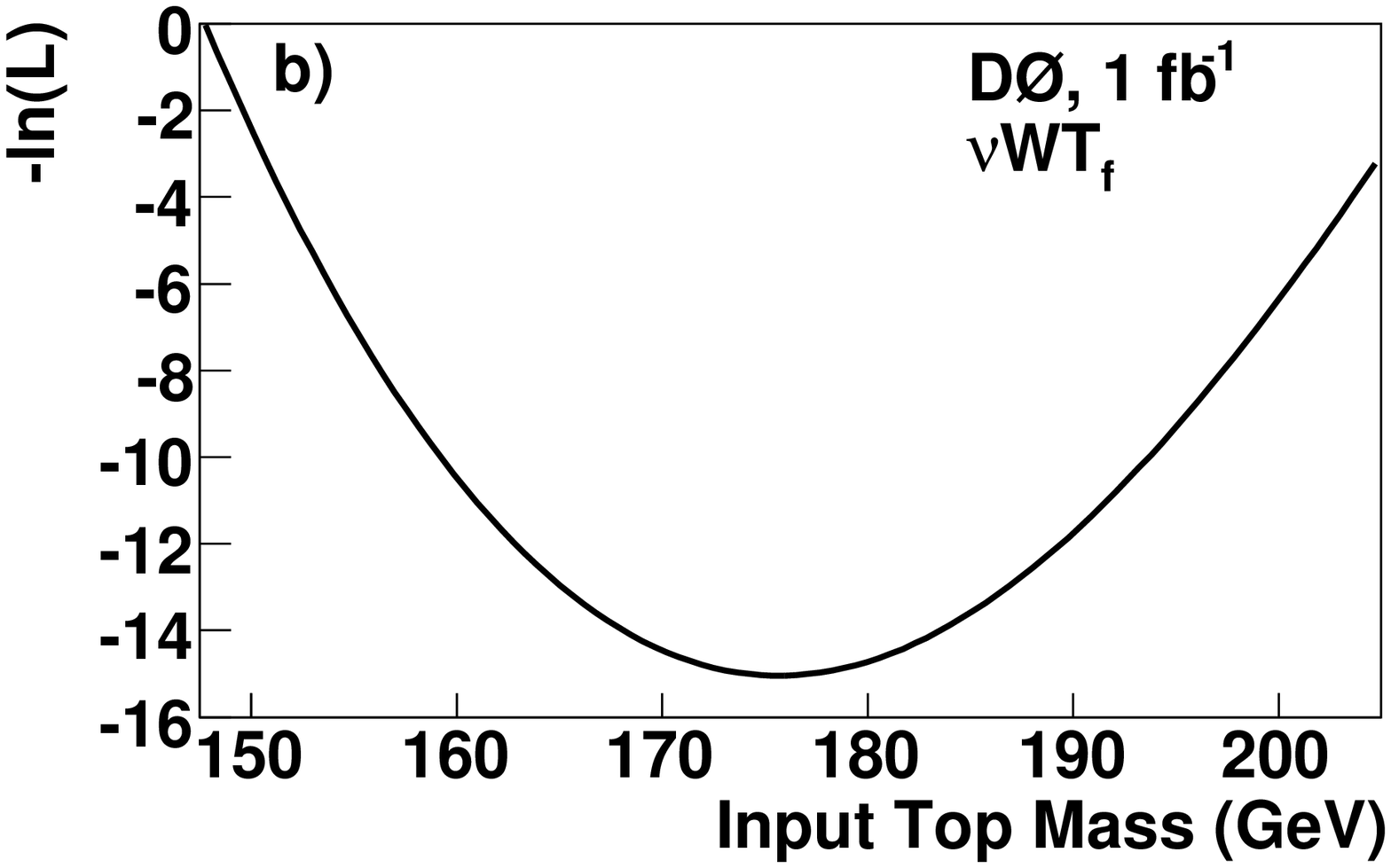}
\includegraphics[scale=0.5]{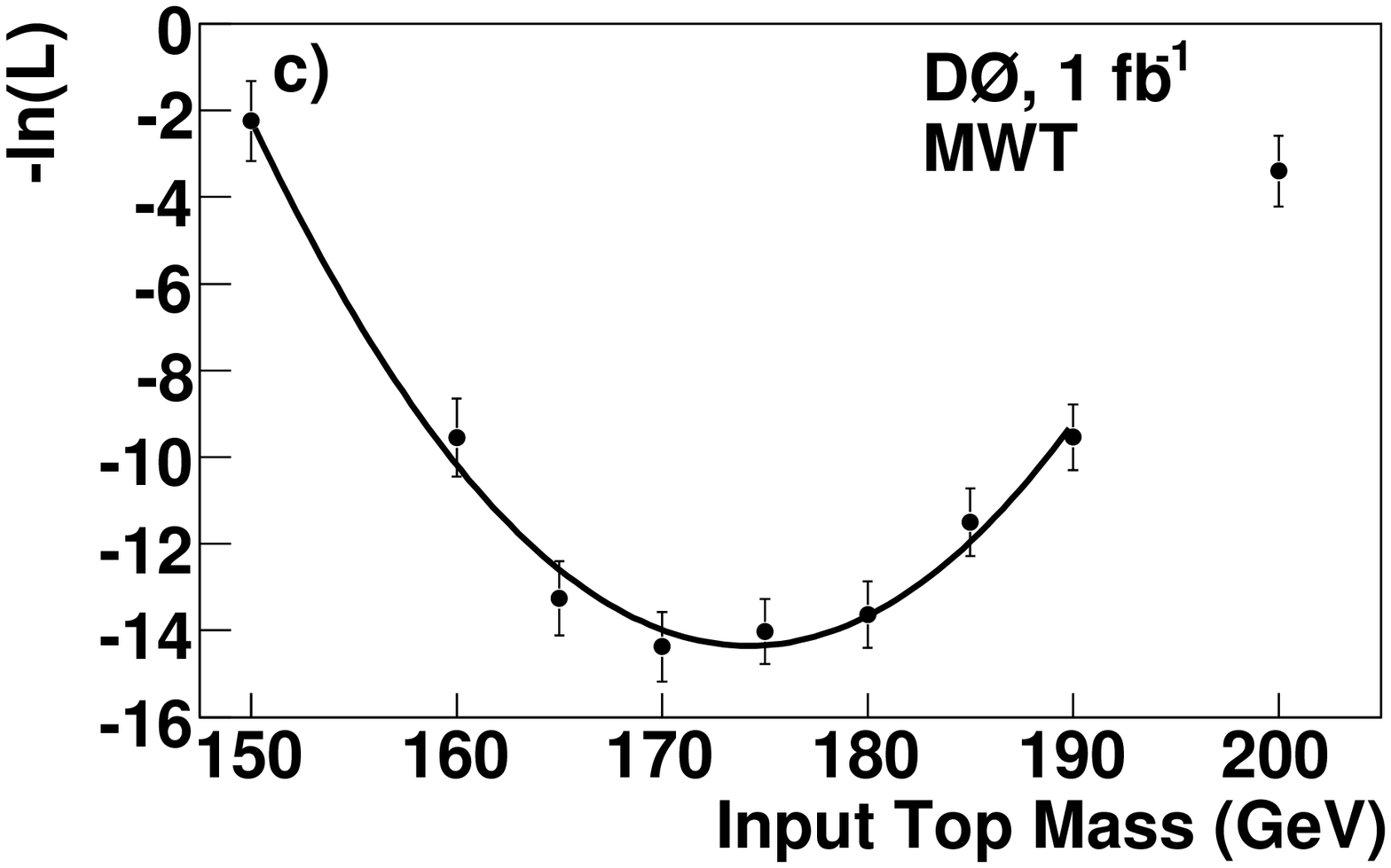}
\caption{Negative log-likelihood $-\ln{\mathcal L}$ vs. $m_t$ 
for the combination of all channels before calibration for the (a) \nuwth, (b) \nuwtf, and (c) MWT methods.}
\label{fig:likeli_combined}
\end{figure*}

{\section{Results}
\label{sec:Results}}
The calibrated mass and statistical
uncertainties for the $2\ell$, $\ell+$track, and their combination are 
shown in Table~\ref{tab:statResult} for each of the three methods.
The $-\ln{\mathcal L}$ fits from the \nuwth, \nuwtf, and MWT methods, 
including data points,
are shown in Fig.~\ref{fig:likeli_combined}. There are no data points for the \nuwtf\ fit 
since the corresponding curve is a one-dimensional slice of an analytic three-dimensional fit function, $f_s$.
The calibrated statistical uncertainties 
determined in data from these likelihood curves
are shown by arrows in Fig.~\ref{fig:staterr_combined}.
The statistical uncertainties agree with the expectations from ensemble testing.
\newline

\begin{table*}
\caption{Calibrated fitted $m_t$ for the \nuwth, \nuwtf, and MWT methods.  
All uncertainties are statistical.}
\begin{center}
\begin{tabular}{cccccccccc}
\hline\hline
Channel 	& \multicolumn{3}{c}{\nuwth\ [GeV]}   & \multicolumn{3}{c}{\nuwtf\ [GeV]} & \multicolumn{3}{c}{MWT [GeV]}\\
\hline
$2\ell$  	& $177.5$&$\pm$&$ 5.5	$ & $176.1$&$\pm$&$5.8$ & $176.6$&$\pm$&$5.5$	 \\
\ltrk	   	& $170.7$&$\pm$&$ 12.3$ & $174.6$&$\pm$&$13.8$  & $165.0$&$\pm$&$8.5$   \\
combined   	& $176.3$&$\pm$&$ 4.9	$ & $176.0$&$\pm$&$5.3$& $173.2$&$\pm$&$4.9$  	\\
\hline
\hline
\end{tabular}
\label{tab:statResult}
\end{center}
\end{table*}

\subsection{Systematic Uncertainties}
\label{systematics}
The top quark mass measurement relies substantially on the Monte Carlo simulation
of \ttbar\ signal and backgrounds. While we have made adjustments
to this model to account for the performance of the detector, residual
uncertainties remain.  The limitations of modeling 
of physics processes may also affect the measured mass.
There are several categories of systematic uncertainties: 
modeling of physics processes, modeling of the detector response, 
and the method.
We have estimated each of these as follows.\\

\subsubsection{\bf Physics Modeling }

\paragraph{$b$ fragmentation} A systematic uncertainty arises from the different models 
of $b$~quark fragmentation, namely the distribution of the fraction of energy taken by the heavy hadron.
The standard \dzero\ simulation used for this analysis utilized the default \pythia\ tune
in the Bowler scheme~\cite{bowler}.  We reweighted our \ttbar\ simulated samples to reach
consistency with the fragmentation model 
measured in $e^{+}e^{-}\to Z\to b\bar{b}$ decays~\cite{bfrag}.
A systematic uncertainty is assessed by comparing the measured $m_t$ in these two scenarios.\\

\paragraph{Underlying event model} An additional systematic uncertainty can arise from the underlying event model. 
We compare measured top quark masses for the \pythia\ tune DW~\cite{DW} with the nominal
model (tune A).\\

\paragraph{Extra jet modeling} Extra jets in top quark events from gluon radiation can affect the \ttbar\
$p_T$ spectrum, and therefore the measured $m_t$.  
While our models describe the data within uncertainties for all channels, the ratio
of the number of events with only two jets to those with three or 
more jets is typically four in the Monte Carlo and three in the data.
To assess the affect of this difference, we reweight the simulated events with a 
top quark mass of 170 
GeV so that this ratio is the same. Pseudoexperiments with 
reweighted events are compared to the nominal pseudoexperiments to determine the uncertainty.\\

\paragraph{Event generator} There is an uncertainty in event kinematics due to 
the choice of an event generator. 
This can lead to an uncertainty in the measured top quark mass.
To account for variations in the accuracy of $t\bar{t}$ generators, we compare
pseudoexperiment results using \ttbar\ events generated with \alpgen\ to those generated with
\pythia\ for $m_t=170$ GeV.  
The difference between the two estimated masses is corrected by subtracting 
the expected statistical uncertainty divided by the 
square root of the number of pseudoexperiments.\\

\paragraph{PDF variations} The top quark mass measurement relies on Monte Carlo events generated 
with a particular PDF set (\cteq). 
Moreover, this PDF set is used directly by the MWT method. We estimate the resulting 
uncertainty on $m_t$ by 
reweighting the Monte Carlo according to the different eigenvectors of the \cteq\ PDFs.  
For each choice, we measure a new mass and the difference between the mass obtained with reweighting and 
a nominal mass is computed. 
The resulting uncertainty is the sum in quadrature of all above uncertainties.\\

\paragraph{Background template shape}  The uncertainties on the background kinematics 
can affect the template shapes and consequently the measured top quark mass.
The uncertainty from the background template shape is found by 
substituting simulated $WW$ events for all $Z$ backgrounds (including $Z\to\tau\tau$) in all pseudoexperiments. 
The uncertainty is taken as the difference between the average measured top quark mass
with this assumption and the nominal value.\\

\subsubsection{\bf Detector Modeling }

\paragraph{Jet energy scale} Because the $b$ jets carry the largest share of the energy in top quark
events, their calibration has the largest affect on the uncertainty
on $m_t$. Ideally, the procedure to calibrate jet energies in data and Monte Carlo 
achieves the same energy scale in both. However, each procedure yields a 
systematic uncertainty.  We estimate the resulting uncertainty by
repeating the pseudoexperiments with simulated events in which the jet energies are 
shifted up and down by the known $p_T$ and $\eta$ dependent uncertainty, obtained by summing in quadrature uncertainties on the data and Monte Carlo jet energy scales.  The probability
density histograms ($h_s$, $h_b$) and functions ($f_s$, $f_b$) 
are left with the nominal calibration.\\

\paragraph{$b$/light quark response ratio} This uncertainty arises from the fact that the jets in signal events 
are primarily $b$ jets. These have a different detector response than
the light quark and gluon jets which dominate the $\gamma+$jet sample 
used to derive the overall jet energy calibration. By applying this calibration to the 
$b$ jet sample, a 1.8\% shift in jet $p_T$ is observed~\cite{ljets_prl}.  We adjust the jets in the
Monte Carlo for this and
propagate the correction into the \met. This causes a shift in the 
measured $m_t$ which is taken as an uncertainty.\\

\paragraph{Sample dependent jet energy scale} After the initial calibration, a residual shift in jet $p_T$ distributions is
observed in $Z+$jets events when comparing data and Monte Carlo.  
We adopt a further calibration that improves agreement in these distributions
and apply it to all of our background samples. 
Because this correction may not apply to \ttbar\ events,
we take the shift in the measured $m_t$ to be a systematic uncertainty.\\

\paragraph{Object resolution} 
The jet resolution from the simulation is better than that observed in data.
To improve the agreement, we apply an additional smearing to Monte Carlo events. A residual
difference between data and Monte Carlo jet resolutions can lead to a mass bias.
To estimate the effect on the $m_t$ measurement, we 
repeat the pseudoexperiments by adjusting this smearing up and down within its uncertainty
while keeping $h_s$ and $h_b$ with the nominal resolutions.
We estimate the systematic uncertainties arising from the muon, isolated track, and electron
$p_T$ resolutions in a similar way.\\

\paragraph{Jet identification} Jet reconstruction and identification efficiency in 
Monte Carlo is corrected
to match the data. We propagate the uncertainty on the correction factor to the 
top quark mass measurement.\\

\paragraph{Monte Carlo corrections} Residual uncertainties on the Monte Carlo corrections exist for 
triggering, luminosity 
profiles, lepton identification, and $b$-tagging. These uncertainties affect the top quark mass 
uncertainties.
In each case, a respective systematic uncertainty on $m_t$ is found by reweighting events according 
to the uncertainties of Monte Carlo correction factors introduced to reproduce the data.\\

\subsubsection{\bf Method }
\label{method}
\paragraph{Background yield} Due to the limited statistics of the simulation, there 
is some uncertainty in the event yields for
the background processes. This uncertainty affects the likelihood and the measured 
top quark mass via Eq.~\ref{eq:Lhisto}. 
To estimate the effect of the uncertainty on background event yield, we vary
the total background yield by its known uncertainty up and down keeping the relative 
ratios of individual background processes constant.\\
  
\paragraph{Template statistics} The templates used in the MWT and \nuwth\ methods have 
finite statistics.
Local fluctuations in these templates can cause local fluctuations in the 
individual likelihood fits and the top quark mass.
We obtain an uncertainty in $m_t$ by
varying the $-\ln{\mathcal L}$ points from the data ensemble within their errors. 
The width of the $m_t$ distribution provides the systematic uncertainty.

For the \nuwtf\ method, the $f_s$ function depends 
on fifteen parameters, each of which
has a corresponding uncertainty. Consequently, there is some uncertainty on the 
shape of this function.  There is a corresponding uncertainty on the parameters of the $f_b$.
The uncertainty on the shape causes the fit uncertainty on the measured top quark mass.
We find the impact of this uncertainty on the data sample by varying the parameters of 
$f_s$ and $f_b$ within their uncertainties.
For each such variation, we remeasure $m_t$ for the data sample. The fit uncertainty 
is the width of this distribution. \\

\paragraph{Monte Carlo calibration} There is an uncertainty on fitting the 
parameters (slope and offset) of the calibration curve.
This uncertainty causes an uncertainty in the calibrated top quark mass.
The calibration uncertainty is obtained as the uncertainty of the offset.\\

\begin{table*}[ht!]
\caption{\label{tab:syst_sum_combination} Summary of systematic uncertainties for the combined analysis of all dilepton channels. The \nuwth, \nuwtf, and MWT method results are shown.}
\begin{center}
\begin{tabular}{lccc}
\hline\hline
Source of uncertainty	& \nuwth\	& \nuwtf\ 	& MWT  \\
			& [GeV]		& [GeV]		& [GeV] \\
\hline
$b$ fragmentation 	& 0.4		& 0.5 		& 0.4		\\
Underlying event modeling& 0.3		& 0.1		& 0.5		\\
Extra jets modeling 	& 0.1		& 0.1		& 0.3		\\
Event generator	 	& 0.6		& 0.8		& 0.5	\\
PDF variation 		& 0.2		& 0.3		& 0.5		\\
Background template shape & 0.4		& 0.3		& 0.3		\\
Jet energy scale (JES) 	& 1.5		& 1.6 		& 1.2	\\
$b$/light response ratio& 0.3		& 0.4 		& 0.6	\\
Sample dependent JES 	& 0.4		& 0.4		& 0.1		\\
Jet energy resolution 	& 0.1		& 0.1		& 0.2	\\
Muon/track $p_T$ resolution & 0.1	& 0.1		& 0.2		\\
Electron energy resolution & 0.1	& 0.2		& 0.2	\\ 
Jet identification	& 0.4		& 0.5		& 0.5		\\
MC corrections	 	& 0.2		& 0.3		& 0.2		\\
Background yield 	& 0.0		& 0.1		& 0.1	\\
Template statistics 	& 0.8		& 1.0		& 0.8		\\
MC calibration		& 0.1		& 0.1		& 0.1		\\
\hline
Total systematic uncertainty 	& 2.1		& 2.3		& 2.0		\\
\hline\hline
\end{tabular}
\end{center} 
\end{table*}

A summary of estimated systematic errors for the combined dilepton
channels is provided in Table~\ref{tab:syst_sum_combination}. 
We assumed the systematic uncertainties for all three methods to be completely correlated for each source of systematic uncertainty and uncorrelated among different sources.
The correlations of statistical uncertainties are given in the next Section.
All uncertainties are corrected for the slope of the mass scale
calibration.  The total uncertainty is
found by assuming all the contributions are independent and adding them in 
quadrature.

\subsection{Combined Results}
\label{combinations}

Because the statistical use of the \nuwt\ moments template is
different between the \nuwth\ and \nuwtf\ methods, 
ensemble tests show that these two 
measurements are only 85\% correlated.
We combine them using the best linear unbiased estimator (BLUE) method~\cite{blue} 
to form a final \nuwt\ measurement.  Applying the correlation
to the measurements from data, we obtain a final \nuwt\ measurement of 
$m_t= 176.2\pm4.8(\rm stat)\pm2.1(\rm sys)$~GeV for the combination of all five channels.
We treat all systematic uncertainties as 100\% correlated across methods except
for the Monte Carlo calibration and template statistics uncertainties.  These
are treated as uncorrelated and 85\% correlated, respectively.
The individual systematic uncertainties on the \nuwt\ combination are the same as 
those for the \nuwth\ method to the precision given in Table~\ref{tab:syst_sum_combination}.  

\begin{table*}
\caption{Final results for the \nuwt\ method and \nuwt +MWT combination.}
\begin{center}
\begin{tabular}{lll}
\hline\hline
Channel 	 & \nuwt\ [GeV] & \nuwt+MWT [GeV]\\
\hline
$2\ell$  	& $177.1\pm\;\;5.4(\rm stat)\pm2.0(\rm sys)$	 & $176.9\pm4.8(\rm stat)\pm1.9(\rm sys)$ \\
\ltrk	   	& $171.2\pm12.3(\rm stat)\pm2.7(\rm sys)$        & $165.7\pm8.4(\rm stat)\pm2.4(\rm sys)$\\
combined   	& $176.2\pm\;\;4.8(\rm stat)\pm2.1(\rm sys)$	 & $174.7\pm4.4(\rm stat)\pm2.0(\rm sys)$\\
\hline
\hline
\end{tabular}
\label{tab:totalResult}
\end{center}
\end{table*}

The final MWT measurement is
$m_t= 173.2\pm4.9(\rm stat)\pm2.0(\rm sys)$~GeV for the combination 
of all dilepton channels.
The total systematic uncertainties are 2.0~GeV and 2.4~GeV for the $2\ell$ and
\ltrk\ MWT results, respectively.  
The \nuwt\ and MWT approaches use partially different
information from each \ttbar\ event; the two results are
measured to be 61\% correlated.  Therefore, we use the BLUE
method to determine an overall measurement of 
$m_t=174.7\pm4.4(\rm stat)\pm2.0(\rm sys)$~GeV for the combination of all dilepton channels.
We treat all systematic uncertainties as 100\% correlated across methods except for
two uncertainties.
The Monte Carlo calibration and template statistics systematic uncertainties
are treated as uncorrelated and 61\% correlated, respectively.
The channel-specific results for both measurement combinations 
are given in Table~\ref{tab:totalResult}.
\newline

{\section{Conclusions}
\label{sec:Conclusions}}

In $1\;\rm fb^{-1}$ of \ppbar\ collision data from the Fermilab Tevatron collider, we employed
two mass extraction methods to measure $m_t$ in \ttbar\ events 
with two high $p_T$ final state leptons.  We analyzed three channels
with two fully identified leptons ($e\mu$, $ee$, and $\mu\mu$) and
two channels with relaxed lepton selection and a $b$-tagged jet ($e+$track
and $\mu+$track).  Using the \nuwt\ event reconstruction, we perform
a maximum likelihood fit to the
first two moments of the resulting distribution of relative weight vs.
$m_t$ to measure
\begin{eqnarray}
\nonumber
\it{m}_t     & = & 176.2 \pm 4.8 (\rm stat) \pm
  2.1 (\rm sys) \,{\rm GeV}.\\
\end{eqnarray}
We also employ the MWT method using a fit to the mass giving the
maximum weight.  We measure
\begin{eqnarray}
\nonumber
\it{m}_t     & = & 173.2 \pm 4.9 (\rm stat) \pm
  2.0 (\rm sys) \,{\rm GeV}.\\
\end{eqnarray}
\noindent Accounting for correlations in these two
measurements, we obtain a final combined result of
\begin{eqnarray}
\nonumber
\it{m}_t     & = & 174.7 \pm 4.4 (\rm stat) \pm
  2.0 (\rm sys) \,{\rm GeV}.\\
\end{eqnarray}
Our result is consistent with the
current world average value of $m_t$ \cite{mtwavg}.

% acknowledgement_paragraph_r2.tex     11/25/08

We thank the staffs at Fermilab and collaborating institutions, 
and 
acknowledge support from the 
DOE and NSF (USA);
CEA and CNRS/IN2P3 
(France);
FASI, Rosatom and RFBR (Russia);
CNPq, FAPERJ, FAPESP and 
FUNDUNESP (Brazil);
DAE and DST (India);
Colciencias (Colombia);
CONACyT 
(Mexico);
KRF and KOSEF (Korea);
CONICET and UBACyT (Argentina);
FOM (The 
Netherlands);
STFC (United Kingdom);
MSMT and GACR (Czech Republic);
CRC 
Program, CFI, NSERC and WestGrid Project (Canada);
BMBF and DFG (Germany);
SFI 
(Ireland);
The Swedish Research Council (Sweden);
CAS and CNSF (China);
and 
the
Alexander von Humboldt Foundation (Germany).

\end{document}